\documentclass[journal,draftclsnofoot,12pt,onecolumn]{IEEEtran}
\usepackage{etex}
\usepackage[a4paper,left=1in,right=1in,top=1in,bottom=1in]{geometry}
\usepackage{textcomp}
\usepackage{cite}
\usepackage{algpseudocode}
\usepackage{pict2e}
\usepackage[table]{xcolor}
\usepackage{supertabular}
\usepackage{booktabs}
\usepackage{multirow}

\usepackage{amsmath}
\usepackage{nccmath}
\usepackage{amssymb} 
\usepackage{amsthm}
\usepackage{bbm}
\usepackage[boxruled]{algorithm2e}
\usepackage{graphicx}
\usepackage[font=footnotesize]{caption}
\usepackage[font=footnotesize]{subcaption}
\usepackage{verbatim}
\usepackage{longtable}
\usepackage{array}
\usepackage{anyfontsize}
\usepackage{bm}
\usepackage{tikz,pgf}
\usepackage{rotating}
\usetikzlibrary{arrows,automata,positioning}
\usepackage{arydshln}
\usepackage{slashbox}	
\usepackage{enumitem}
\setlist[itemize]{leftmargin=*}
\graphicspath{{Figs/}}
\theoremstyle{plain} 
\theoremstyle{plain} 
\theoremstyle{definition} 
\theoremstyle{definition} 
\theoremstyle{definition} 
\theoremstyle{remark} 
\theoremstyle{plain} 

\makeatletter
\def\bbordermatrix#1{\begingroup \m@th
	\@tempdima 4.75\p@
	\setbox\z@\vbox{%
		\def\cr{\crcr\noalign{\kern2\p@\global\let\cr\endline}}%
		\ialign{$##$\hfil\kern2\p@\kern\@tempdima&\thinspace\hfil$##$\hfil
			&&\quad\hfil$##$\hfil\crcr
			\omit\strut\hfil\crcr\noalign{\kern-\baselineskip}%
			#1\crcr\omit\strut\cr}}%
	\setbox\tw@\vbox{\unvcopy\z@\global\setbox\@ne\lastbox}%
	\setbox\tw@\hbox{\unhbox\@ne\unskip\global\setbox\@ne\lastbox}%
	\setbox\tw@\hbox{$\kern\wd\@ne\kern-\@tempdima\left[\kern-\wd\@ne
		\global\setbox\@ne\vbox{\box\@ne\kern2\p@}%
		\vcenter{\kern-\ht\@ne\unvbox\z@\kern-\baselineskip}\,\right]$}%
	\null\;\vbox{\kern\ht\@ne\box\tw@}\endgroup}
\makeatother
\IEEEoverridecommandlockouts
\begin{document}
\font\myfont=cmr12 at 10pt
\font\myfontTwo=cmr12 at 9pt
\thispagestyle{empty}

\title{Energy Efficiency Through Joint Routing and Function Placement in Different Modes of SDN/NFV Networks}

\author{{\myfont\IEEEauthorblockN{Reza Moosavi\IEEEauthorrefmark{1}, Saeedeh Parsaeefard\IEEEauthorrefmark{2}, Mohammad Ali Maddah-Ali\IEEEauthorrefmark{3}, Vahid Shah-Mansouri\IEEEauthorrefmark{1},  Babak Hossein Khalaj\IEEEauthorrefmark{3}, and , Mehdi Bennis\IEEEauthorrefmark{4}}}\\
{\myfontTwo\IEEEauthorblockA{\IEEEauthorrefmark{1}School of Electrical and Computer Engineering, College of Engineering, University of Tehran, Tehran, Iran}\\
\IEEEauthorblockA{\IEEEauthorrefmark{2}Department of Electrical and Computer Engineering, University of Toronto, Toronto, ON, Canada}\\
\IEEEauthorblockA{\IEEEauthorrefmark{3} Department of Electrical Engineering, Sharif University of Technology, Tehran, Iran}\\
\IEEEauthorblockA{\IEEEauthorrefmark{4}Centre for Wireless Communications, University	of Oulu, 90014 Oulu, Finland}}

{\myfont Email: rmoosavi@ut.ac.ir; saeideh.fard@utoronto.ca; maddah\_ali@sharif.edu; vmansouri@ut.ac.ir; khalaj@sharif.edu; mehdi.bennis@oulu.fi
}}

%\markboth{Journal of \LaTeX\ Class Files,~Vol.~14, No.~8, August~2015}%{Shell \MakeLowercase{\textit{et al.}}: Bare Demo of IEEEtran.cls for IEEE Transactions on Magnetics Journals}
\maketitle

\vspace{-0.5in}
\begin{abstract}
	Network function virtualization (NFV) and software defined networking (SDN) are  two promising technologies to enable 5G and 6G services and achieve cost reduction, network scalability, and deployment flexibility.
	However, migration to full SDN/NFV networks in order to serve these services is a time consuming process and costly for mobile operators. This paper focuses on energy efficiency during the transition of mobile core networks (MCN) to full SDN/NFV networks, and explores how energy efficiency can be addressed during such migration. We propose a general system model containing a combination of legacy nodes and links, in addition to newly introduced NFV and SDN nodes. We refer to this system model as partial SDN and hybrid NFV MCN which can cover different modes of SDN and NFV implementations. Based on this framework, we formulate energy efficiency by considering joint routing and function placement in the network. Since this problem belongs to the class of non-linear integer programming problems, to solve it efficiently, we present a modified Viterbi algorithm (MVA) based on multi-stage graph modeling and a modified Dijkstra's algorithm.
	We simulate this algorithm for a number of network scenarios with different fractions of NFV and SDN nodes, and evaluate how much energy can be saved through such  transition.
	Simulation results confirm the expected performance of the algorithm which saves up to 70\% energy compared to  network where all nodes are always on.
	Interestingly, the amount of energy saved by the proposed algorithm in the case of hybrid NFV and partial SDN networks can reach up to 60-90\% of the saved energy in full NFV/SDN networks.
\end{abstract}
\vspace{-0.3in}
\begin{IEEEkeywords}
	\vspace{-0.1in}
	5G and 6G, Energy efficiency, network function virtualization, software defined networking.
	\vspace{-0.1in}
\end{IEEEkeywords}

%%%%%%%%%%%%%%%%%%%%%%
%%%%%%%%%%%%%%%%%%%%%%
\section{Introduction}
%%%%%%%%%%%%%%%%%%%%%%
%%%%%%%%%%%%%%%%%%%%%%

\label{sec:introduction}
The increase in the number of mobile devices, new services, and applications is leading to an exponential rise in the data traffic in mobile networks. In order to handle such traffic, improve deployment flexibility, and reduce the costs of the service, the fifth-generation mobile networks and beyond (5G and 6G) should rely on new concepts and architectures \cite{AccessManagement}.
Besides, energy consumption in mobile networks increases by 10\% annually and should be appropriately handled \cite{bolla2015finegrained}. The use of software-defined networking (SDN) and network function virtualization (NFV) in the core of 5G can satisfy the new services' demands in the mobile core network (MCN), as well as considerably increase energy efficiency.

NFV leverages the concept of virtualization to network functions (NFs) in which a software implementation of NFs is decoupled from the hardware infrastructure. In this context, virtualized NFs (VNFs) can be installed in/removed from a server, or migrate from one server to another in a straightforward manner \cite{AccessManagement}. This flexibility in NFV enables the migration of VNFs from under-utilized hardware in order to turn them off and achieve higher energy efficiency. On the other hand, SDN  separates the data and control planes and carries out all the control decisions in the network with centralized controllers called SDN controllers.
By a centralized control view, SDN provides a more efficient network control for the operators to attain higher network utilization   \cite{AccessManagement}. For instance, from an energy efficiency perspective, SDN controllers can reroute network paths in such a way that under-utilized links and nodes can be turned off. Inherent potentials of these two complementary technologies in increasing the energy efficiency of the networks can be deployed using the concepts of multi-domain and recursive orchestration in 5G \cite{5g2016view}.

The architecture of MCN of 5G and 6G can be evolutionary, in a way that part of the legacy MCN entities are virtualized or software-defined enabled, while a subset of internal functionalities and existing interfaces remain intact \cite{Corenetworksurvey}. This way, the running cost of MCN of 5G and 6G reduces. Thus, evolutionary architecture for MCN of 5G and 6G, e.g. \cite{115fme,mano5g}, is backward compatible. Motivated by the aforementioned observations, in this paper, an evolutionary architecture for MCN of 5G and 6G is considered as follows:
\begin{itemize}
	\item The transmission network is assumed to be partial SDN, in which some part of the nodes in the network are equipped with SDN technology, while the rest of the nodes remain working independently.
	\item NFs of the data plane are executed in the form of hybrid NFV. This way, some NFs are executed on servers, and some of them are executed on physical function nodes.
\end{itemize}

This hybrid deployment can also model \emph{transition} of the cellular networks from fourth-generation (4G) to 5G and 6G. Mainly because the transition budgets are limited, and only a part of the network can be upgraded at a time, especially for large-scale networks. In addition, as an emerging technology, SDN is not mature enough to replace the whole traditional network in a single step. Therefore, gradual software-defined and  MCN implementation sounds inevitable to assess SDN  feasibility in next generation of wireless networks. Evidently, the transition time is likely to extend over a several years span. During the transition, SDN should coexist with traditional networks \cite{Corenetworksurvey}.  In the partial SDN network, we can still implement centralized control of these SDN-enabled devices to save energy. However, the hybrid SDN is not as effective and flexible as in full SDN.
Similarly, due to the implementation problems and costs, operators seem to be imperceptibly moving towards full NFV networks. Such arguments have led to works such as \cite{115fme} and \cite{110lightweight} to consider hybrid NFV architectures for 5G and 6G MCN.

\begin{figure}
	\vspace{-0.5in}
	\begin{center}
		\includegraphics[width=0.6\textwidth]{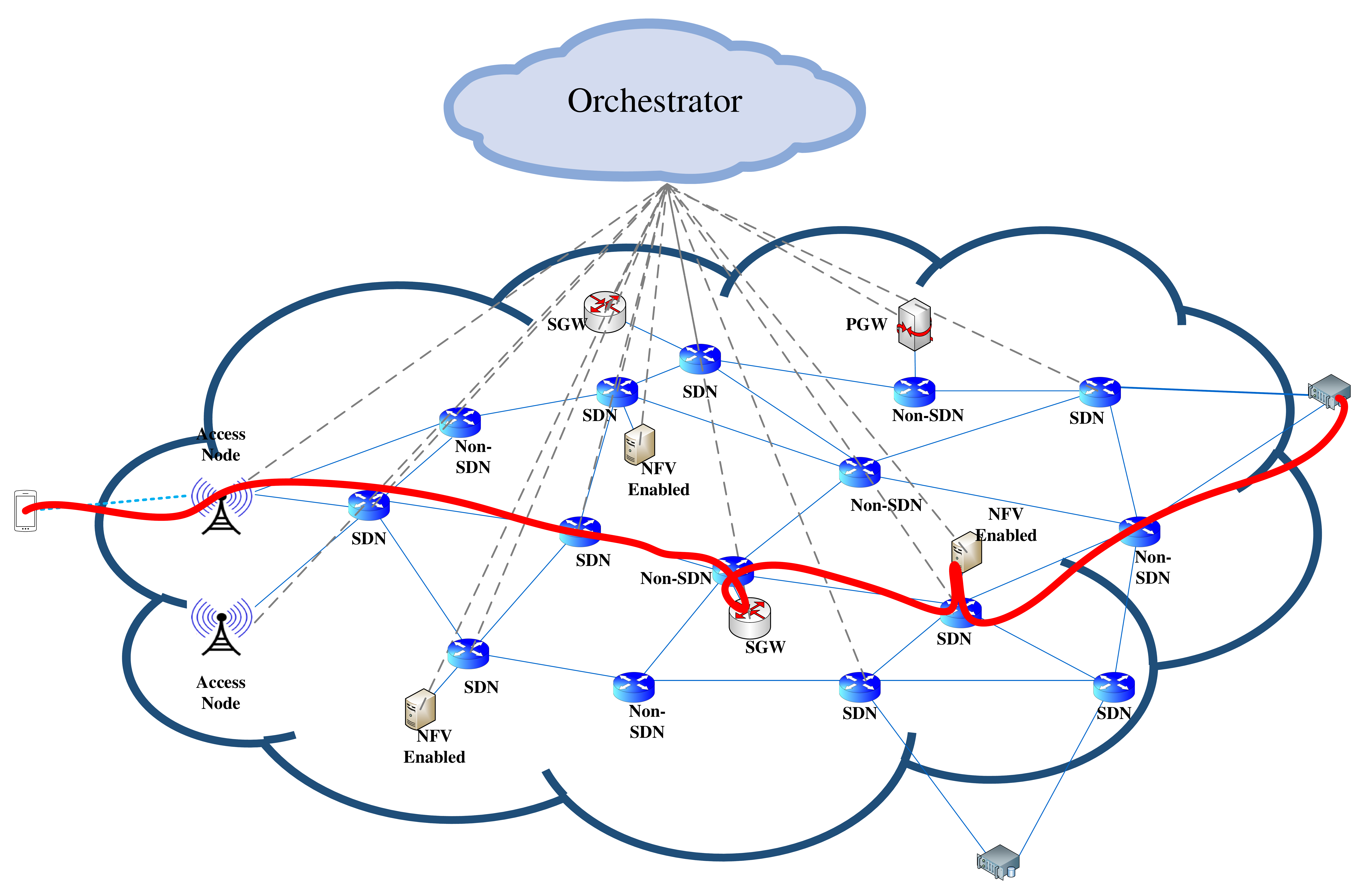}
	\end{center}
	%\vspace{-0.5in}
	\renewcommand{\captionfont}{\small}
	\caption{Partially SDN and hybrid NFV mobile core Network of wireless networks}
	\label{fig1}
	\vspace{-0.3in}
\end{figure}

In this paper, we consider a partial SDN and hybrid NFV  network, as shown in Fig. \ref{fig1} which can model different modes of combinations of SDN and NFV with traditional networking during this transition. In this network, we only consider the data plane where NFs are implemented as hybrid NFV (on servers or physical function nodes), and the transport network is partial SDN. We assume that an orchestrator performs all functions of control plane and control tasks related to NFV and SDN. The proposed system model is comprehensive for SDN/NFV networks where it can be deployed for full/partial SDN networks, full/hybrid NFV networks, or (partial) SDN/(hybrid) NFV networks.

For this setup, we formulate the problem of energy efficiency of the MCN of 5G, and propose an algorithm to save energy in the network. Based on the proposed algorithm, the unused nodes and links in the network are turned off to consume less energy. Our  contributions are summarized as follows:

\begin{itemize}
	\item This paper provides a framework to model the MCN of 5G during transition from 4G to 5G.
	\item The presented work is the first to address the problem of saving energy in partial SDN and hybrid NFV MCN through the transition to 5G. Although energy efficiency, in traditional networks, SDN networks, and partial SDN networks is a well-studied problem, saving energy in partial SDN and hybrid NFV is a new problem that has  not been addressed to the best of our knowledge.
	\item  Energy efficiency in SDN/NFV networks is comprehensively modeled, including different modes of SDN/NFV networks, e.g., (partial)SDN, (hybrid)NFV, partial SDN/hybrid NFV, partial SDN/full NFV, and full SDN/full NFV networks.
	\item The problem of energy efficiency in the considered network is formulated as a non-linear integer programming (NLIP). The objective is to minimize the power consumption of the network components by turning off the unused or less-loaded components.
	\item Because of the NP-hardness of the resulting optimization problem, a suboptimal algorithm based on the Viterbi Algorithm (VA) is proposed to solve the optimization problem. The algorithm starts via assigning weights to all nodes and links based on their on/off-states, as well as their energy consumption levels. The network is modeled as a multi-stage graph, and a candidate set determines the paths of flows, the type of NFs, and their locations to run for each flow.  Ultimately, paths are selected based on a  modified Djikstra's routing algorithm (MDRA). We refer to this algorithm as modified Viterbi algorithm (MVA).
	\item Simulation results demonstrate that the performance of the network in terms of energy efficiency, can be very close to optimum without a full migration to a SDN-NFV platform.
\end{itemize}

The rest of the paper is organized as follows. Related work is presented in Section \ref{related}. In Section \ref{system_model}, the system model is described. In Section \ref{problem}, the problem formulation of energy efficiency in partial SDN and hybrid NFV  MCN for 5G is introduced. Subsequently, in Section \ref{algorithmsection}, an efficient algorithm based on MVA  is proposed and its computational complexity are presented. In Section \ref{simulation}, the performance of our proposed algorithm is evaluated for different network sizes. Finally, Section \ref{conclusion} concludes the paper.

%%%%%%%%%%%%%%%Related Work%%%%%%%%%%%%%%
\section{Related Work}\label{related}
This paper sits in the intersection of three groups of works: research on the energy efficiency in SDN networks; the energy efficiency in the partial SDN networks and  efficiency in the NFV.
We will review these three of reserach trends and then we will highlight the novelties of this paper.

For full SDN networks, there exists a large body of works focusing on the energy efficiency either in data center networks, e.g., \cite{gal13openflow,partially17Energyefficientdatacenter}, or on the conventional data networks. e.g., \cite{gal18Energyefficientminimumspanningtree}. For the former cases, usually under-utilized data centers are switched off via SDN controllers\cite{gal13openflow,partially17Energyefficientdatacenter}. However, for the latter case, by considering the physical structure of switches, heuristic algorithms are proposed to reduce the energy consumption \cite{linecards,rule2014optimizing}. None of these two cases is related to MCN and to fill this gap, we study the energy efficiency for the MCN armed by SDN and NFV for different scenarios of implementation e.g., partial SDN and hybrid NFV.

In \cite{partially}, the problem of energy saving in partial SDN networks is addressed  by switching off the SDN links and nodes.\footnote{Partial SDN networks are first mathematically introduced in \cite{agarwal2013trafficeng} where a fraction of switches are equipped with SDN capability..} Those links and nodes are chosen through optimizing a novel objective function using a heuristic algorithm. In \cite{bolla2015finegrained}, a partial SDN is deployed via a green abstraction layer which belongs to the European Telecommunications Standards Institute (ETSI) \cite{gal}. In this standard, multi energy-aware states are considered for the devices. In each state, the device consumes a certain amount of energy and attains a specific performance. An energy efficiency metric named Ratio for Energy Saving in SDN (RESDN) that quantifies energy efficiency based on link utility intervals is presented by \cite{RESDN}. A heuristics algorithm for maximize RESDN is presented.

NFV placement problems have received a lot of interest recently \cite{MAT,Behrooz}.
\cite{kerim,boutaba,ammar2017migration} apply the Viterbi algorithm to solve their NFV resource allocation problems.
In \cite{ammar2017migration,ammar_acm}  where NFVs are allocated to servers in the network to optimize objective functions, e.g., energy efficiency, subject to network limitations and required quality of services. These problems are  inherently high dimensional and complex problems \cite{resource}.  For instance, \cite{ammar_acm}  deploys a Markov decision process  to minimize energy consumption in NFV based networks. \cite{boutaba}  has considered four types of costs for the problem of resource allocation in NFV, one of which is energy resources.

\cite{TGCNshojafar2018joint} also uses migration to decrease energy consumption in the cloud data centers where it models the energy consumption of cloud network by considering the computing costs of the virtual elements on the physical servers, the migration cost for virtual elements across the servers, and the costs of transferring data between the virtual elements. In addition, it introduces a weight parameter to avoid  excessive load of virtual element migrations.
\cite{TNSMkar2017energy} designs a dynamic energy-saving model with NFV technology using an M/M/c queuing network with the minimum capacity policy where a certain amount of load is required to start the virtual machine, and an energy-cost optimization problem is formulated with capacity and delay as constraints.
In \cite{ESSO}, a new energy smart service function chain orchestrator is proposed in which NFV placement is adjusted to utilize more renewable energy in telecommunication networks.
A resource allocation problem, considering constraints on delay, link utilization, and server utilization, is presented in \cite{TNSMtajiki2018joint}, which enables energy-aware SFC for SDN-based networks.
In order to achieve  the trade-off among energy saving, bandwidth usage minimization, and migration cost reduction, \cite{Access2019virtualized} proposes a VNF consolidation method (VCMM), which determines appropriate servers to be turned off by leveraging neural network with multiple characteristics of network status  as input.  VCMM migrates VNFs on servers to turn them off by adopting a greedy mechanism.
%Evidently non of the mentioned recent works considers the hybrid system model of SDN and NFV based networking. In this paper we introduce a general system model to fulfill this gap.
In \cite{mano5g}, an evolutionary architecture based NFV for 5G core is considered and  an energy efficiency problem for data plane and control plane of the architecture is formulated. It also tests the proposed algorithm on a real-world testbed based on OpenStack and OpenDaylight.
\cite{demandResponse} addresses energy efficiency in NFV/SDN for internet of things networks where a new incentive mechanism exploratory works among energy suppliers and consumers is designed in order to encourage consumers to adjust their energy consumption based on available resources.
In this paper, we propose a general system model which can be utilized for any type of full, partial, and non SDN/NFV networks. Saving energy in partial SDN and hybrid NFV MCNs is a new problem that has not been addressed to the best of our knowledge. The presented model and energy efficiency formulation for MCN networks can formulate the transition from 4G to 5G.

%%%%%%%%%%%%%%%%%%%%%System Model%%%%%%%%%%%%%%%%%%%
\section{System Model}\label{system_model}
We consider a partial SDN and hybrid NFV network, represented by an undirected graph $\mathcal{A}=\{\mathcal{N},\mathcal{E}\}$, where $\mathcal{N}=\{1,2,\cdots,N\}$ denotes the set of all nodes including SDN, NFV, and legacy nodes, and  $\mathcal{E}=\{ 1,2,\cdots,E\}$ denotes the set of $E$ undirected links connecting the nodes. In this setup, we assume that a central orchestrator is responsible for providing coordination among all domain controllers, i.e., an SDN controller and the management layers of the legacy nodes. Each mobile traffic flow is specified by its own rate limit and a specific set of NFs, which should be run on this flow in a specific order, called the service chain of this flow. In order to establish an end to end connection for this flow, a specific route to transfer data between the assigned nodes should be determined. In what follows, we provide a mathematical model to represent network limitations and flow characteristics.

\subsection{Network Function Structure}
In this setup, the set of all NFs are represented by $\mathcal{G}=\{g_1,g_2,\cdots,g_K \}$ where each service chain of a flow is a specific subset of $\mathcal{G}$ with predetermined order introduced by the orchestrator.  All these NFs can be run in a virtualized manner,  referred to VNFs in this case. $g_k$ can be VNF $k$ or implemented in a physical function node permanently as an NF.

When $g_k$ is a VNF, it requires a set of resources $\mathcal{C}_{g_k}=\{ c_{k}^{1},c_{k}^{2},\cdots,c_{k}^{L} \}$ where $c_k^l$ represents the required resource of category $l$ for VNF $g_k$. In addition, the amount of ingress traffic rate that $g_k$ in a virtualized manner can process is $r_k$. We refer to $r_k$ as a processing capacity  of VNF $g_k$ \cite{boutaba}.

\subsection{Flow Representation}
We assume that flow arrivals occur in a time-slotted manner, and the flows are received during a time-slot to be served at the beginning of the following time-slot. Consider a set of flows in a network as $\mathcal{F}$. Each mobile traffic flow $f$ can be represented as $f=(s^{f},d^{f},r^{f},\mathcal{G}^{f})$ where $s^{f}$, $d^{f}$ and $r^{f}$ denote the source, the destination, and the amount of incoming rate of flow $f \in \mathcal{F}$, respectively. Here, we show a service chain of flow $f$ as  $\mathcal{G}^{f}=\{g_{1}^{f},g_{2}^{f},\cdots,g_{\bar{K_f}}^{f}\}$, which is a set of NFs with a specific order to  be run over flow $f$. $\bar{K_f}$ denotes the number of NFs running over flow $f$, and $g_{\bar{k}}^{f}$ denotes $\bar{k}^{\text{th}}$ NF that must be run for flow $f$. Note that $g_{\bar{k}}^{f}$ also belongs to $\mathcal{G}$, however to show the order of NFs for each flow and their related sequences, we use this notation per flow. The orchestrator has knowledge about which $g_k$ is related to $g^f_{\bar{k}}.$
In this paper, to denote the mapping of $g_k$ to $g_{\bar{k}}^{f}$ via Orchestrator, we use the following notation

\begin{equation*}\label{mapping}
	\chi^f_{k\rightarrow \bar{k}}(g_k)=g^f_{\bar{k}} \quad \forall g_k\in \mathcal{G} , ~\forall g^f_{\bar{k}}\in \mathcal{G}^f, ~ \forall f,
\end{equation*}
where $\chi^f_{k\rightarrow \bar{k}}(g_k)$ represents the mapping stored in the orchestrator and utilized during the service time of flow $f$.

\subsection{Nodes' Categories and Features}
In the mixed structure of traditional SDN and NFV-based networks in our setup,  to run a service chain of any user's request and provide the connections,  there exist four sets of network nodes, i.e.,
\begin{itemize}
	\item $\mathcal{N_{T}}=\{ 1,2,\cdots,N_T\}$ as the set of traditional switches, called  non-SDN nodes;
	\item $\mathcal{N_{S}}=\{ 1,2,\cdots,N_S\}$ as the set of SDN enabled switches, called  SDN nodes;
	\item $\mathcal{N_{N}}=\{ 1,2,\cdots,N_N\}$ as the set of NFV based servers running  NFs in a virtualized manner and is referred  to as NFV nodes;
	\item $\mathcal{N_{M}}=\{ 1,2,\cdots,N_M\}$ as the set of physical function nodes, such as nodes of serving gateways (SGW) and packet data network gateways (PGW) and is called non-NFV nodes.
\end{itemize}
As a result, the total number of nodes in the network is $N_T+N_S+N_N+N_M=N$ where  each node $u\in \mathcal{N}$ has two states: 1) \textit{On-state} in which $u$ is on; and, 2) \textit{Off-state} where $u$ is off. The orchestrator has  access to the states of all nodes however, the state of SDN/NFV nodes, eg.,  $\mathcal{N} \backslash \mathcal{N_{T}}$ can be dynamically adjusted.  To represent states of the nodes, we define $\alpha (u)$ as

\begin{equation*}
	\alpha (u)=
	\left\{\begin{array}{ll}
		0, & $if $ u \in \mathcal{N}\backslash\mathcal{N_{T}}$ is in off-state,$ \\
		1, & $otherwise$.
	\end{array}\right.
\end{equation*}

To run a set $\mathcal{G}^f$ for each flow $f\in \mathcal{F}$, orchestrator requires information about NFs that each $u\in \mathcal{N_{M}}\cup \mathcal{N_{N}}$ can run, and the amount of capacity of each $u\in \mathcal{N_{M}}\cup \mathcal{N_{N}}$.
Each non-NFV node $u\in \mathcal{N_{M}}$ can run a predetermined set of NFs.
Let \textbf{$\mathcal{G}_{u}=\{g_{u_{1}},g_{u_{2}},\cdots,g_{u_{k_{u}}}\}$} represent the NFs which are run in node $u\in \mathcal{N_{M}}$. Each node $u\in \mathcal{N_{M}}$  has limited capacity of ingress rate denoted by $r_{u}$.

Each NFV node $u\in \mathcal{N_{N}}$ can run all NFs in $\mathcal{G}$, and has a limited capacity for  $L$ types of required resources of  NF. This limitation  is denoted by $\mathcal{C}_{u}=\{c_{u}^{1},c_{u}^{2},\cdots,c_{u}^{L}\}$ where $c_u^l$ is the maximum capacity of node $u$ for  type $l$ of resources. At any instance, each  node $u\in \mathcal{N_{N}}$ can just run one instance of VNF $k$.
Specifically, if  NF $g_k$  is run over $u\in \mathcal{N_{N}}\cup \mathcal{N_{M}}$, we define
\begin{equation*}
	\mu_{u}^{k}=
	\left\{\begin{array}{ll}
		1, & $if $ $node $u\in \mathcal{N_{N}}\cup \mathcal{N_{M}}$ runs  $g_k, \\
		0, & $otherwise.$
	\end{array}\right.
\end{equation*}
To demonstrate that flow $f$ is served by node $u$ to run $g_{\bar{k}}$, we define

\begin{equation*}
	\mu_{u}^{\bar{k}f}=
	\left\{\begin{array}{rl}
		1, & $if $ $node $u$ is selected to run $g_{\bar{k}}^f$ of flow $f, \\
		0, & $otherwise.$
	\end{array}\right.
\end{equation*}

One practical point is that there exists a situation that more than two consecutive NFs are run in one node $u \in \mathcal{N_{M}}$ for flow $f$. In this situation the incoming rate to non-NFV node $u\in \mathcal{N_{M}}$ is equal to the rate of the first NF of flow $f$ deployed on $u\in \mathcal{N_{M}}$ \cite{gsmlte}. To consider this point in our formulation, for all $u\in \mathcal{N_{M}}$, we define the following variable
\begin{equation*}
	\zeta_{u}^{\bar{k}f}=
	\left\{\begin{array}{rl}
		0, & $if $\mu_{u}^{\bar{k}f}\mu_{u}^{\bar{k}-1f}=1, \\
		1, & $otherwise.$
	\end{array}\right.
\end{equation*}

\subsection{ Notations for Links and Paths}
In our system model, $(u,v)\in \mathcal{E}$ is a physical connection between two nodes $u\in \mathcal{N}$ and $v \in \mathcal{N}$. Each node $u\in \mathcal{N_{M}} \cup \mathcal{N_{N}}$ only connects to a switch with a link, and the switch connects to other switches with one or more links.
For each link $(u,v) \in \mathcal{E}$, $c(u,v)$ denotes its capacity. When $u$ or $v\in \mathcal{N}\backslash\mathcal{N_{T}}$, link $(u,v)$ is referred to as SDN link, and a set of all SDN links is denoted by $\mathcal{E_{S}}\subseteq \mathcal{E}$. In our setup, for each SDN link, orchestrator considers two states: 1) \textit{On-state} if link $(u,v)\in\mathcal{E_{S}}$ is on; 2) \textit{Off-state} if $(u,v)\in\mathcal{E_{S}}$ is off. To denote the on and off-states of link $(u,v)\in \mathcal{E_{S}}$, we define
\begin{equation*}
	\beta(u,v)=
	\left\{\begin{array}{rl}
		1, & $if$~ (u,v)\in \mathcal{E_{S}} ~$is active,$ \\
		0, & $otherwise.$
	\end{array}\right.
\end{equation*}
Note that the orchestrator can control on-states and off-states of all SDN links.

When a physical connection between two nodes $u$ and $v$ does not exist, we consider a set of paths between two nodes denoted by $\mathcal{P}_{u\rightarrow v}= \{ \mathcal{P}^{1}_{u \rightarrow v},\mathcal{P}^{2}_{u \rightarrow v},\cdots,\mathcal{P}_{u \rightarrow v}^{J_{u \rightarrow v}} \}$, where $J_{u \rightarrow v}$ denotes number of all paths between nodes $u$ and $v$. To mathematically represent which path   $\mathcal{P}^{j}_{u \rightarrow v}$ is selected  for the flow $f$, we define
\begingroup
\begin{align*}
	\Gamma_{u \rightarrow v}^{fj}=
	\left\{\begin{array}{rl}\textbf{}
		1, & \text{\small if }\mathcal{P}^{j}_{u \rightarrow v}\text{\small is selected for flow }f~\text{\small between nodes }u~\text{\small and } v , \\
		0, & \text{\small otherwise,}
	\end{array}\right.
\end{align*}
\endgroup
and, to show whether or not physical connection $(u',v')$ belongs to path $\mathcal{P}^{j}_{u \rightarrow v}$, we define
\begingroup
\begin{align*}
	\rho_{u \rightarrow v}^{j(u',v')}=
	\left\{\begin{array}{rl}
		1, & \text{\small if physical connection }(u',v')~\text{\small belongs to }\mathcal{P}^{j}_{u \rightarrow v}, \\
		0, & \text{\small otherwise.}
	\end{array}\right.
\end{align*}
\endgroup

If two different nodes $u$ and $v$ run $g_{\bar{k}}^f$ and $g_{\bar{k}+1}^f$, a path from $\mathcal{P}_{u\rightarrow v}$ is selected for flow $f$. For the case that  one node $u\in \mathcal{N_{M}} \cup \mathcal{N_{N}}$  are run $g_{\bar{k}}^f$ and $g_{\bar{k}+1}^f$, we denote a communication path between this two  NFs by  $\mathcal{P}_{u\rightarrow u}= \{(u,u)\}$ with infinite capacity.

\subsection{Energy Model of Nodes and Links}

When each node $u\in \mathcal{N_{T}} \cup \mathcal{N_{S}}$  is in  on-state, its consumed power is almost constant. Here, we assume that this power is equal to a maximum consumed power of node $u\in \mathcal{N_{T}} \cup \mathcal{N_{S}}$, i.e., $P_u^{\text{max}}$ \cite{partially}. For NFV and non-NFV nodes, a linear model of  power consumption is assumed for on-states \cite{measurement}. For instance,  when the ingress rate of node $u\in \mathcal{N_{N}}\cup \mathcal{N_{M}}$  increases, the consumed power increases accordingly. In this paper, we consider following expression for power consumption of a node $u\in \mathcal{N}\backslash\mathcal{N_T}$
% \begingroup\makeatletter\def\f@size{8}\check@mathfonts
%     \begin{align*}
%         a(u)=
%         \left\{\begin{array}{ll}
%         P_{u}^{\text{max}}, &\text{\small node }u\in \mathcal{N_S}~\text{\small is in on-state},\\
%         (\theta + (1-\theta)\frac{r^{c}_{u}}{r_{u}})P_{u}^{\text{max}}, &\text{\small node }u\in \mathcal{N_N}\cup \mathcal{N_M} ~\text{\small is in on-state},\\
%         0, &\text{\small node }u\in \mathcal{N}\backslash\mathcal{N_T} ~\text{\small is in off-state},
%         \end{array}\right.
%     \end{align*}
% \endgroup
%\begingroup\makeatletter\def\f@size{8}\check@mathfonts
%\begin{align*}
	a(u)=
	%\left\{\begin{array}{ll}
		%P_{u}^{\text{max}},                                             & u\in %\mathcal{N_S}~\text{\small is in on-state},                        \\
		%(\theta + (1-\theta)\frac{r^{c}_{u}}{r_{u}})P_{u}^{\text{max}}, & u\in %\mathcal{N_N}\cup \mathcal{N_M} ~\text{\small is in on-state},     \\
		%0,                                                              & u\in %\mathcal{N}\backslash\mathcal{N_T} ~\text{\small is in off-state},
%	\end{array}\right.
%\end{align*}
%\endgroup
\begingroup
\begin{align*}
	a(u)=
	\left\{\begin{array}{ll}
		P_{u}^{\text{max}},                                             & u\in \mathcal{N_S}~\text{\small is in on-state},                        \\
		(\theta + (1-\theta)\frac{r^{c}_{u}}{r_{u}})P_{u}^{\text{max}}, & u\in \mathcal{N_N}\cup \mathcal{N_M} ~\text{\small is in on-state},     \\
		0,                                                              & u\in \mathcal{N}\backslash\mathcal{N_T} ~\text{\small is in off-state},
	\end{array}\right.
\end{align*}
\endgroup
where $P_{u}^{\text{max}}$ is the maximum power consumed in node $u$ and $\theta$ is a ratio of idle state power to full load power. An idle state represents the state that $u$ is in on-state, however, there is no incoming flow for $u$. $r^{c}_{u}$ and $r_{u}$  denote the amount of current ingress traffic rate and  maximum ingress traffic rate of the node $u\in\mathcal{N_N}\cup \mathcal{N_M} $, respectively.

Consumed power for link $(u,v)\in \mathcal{E}$ is constant in on-state \cite{partially}. When  $(u,v)\in \mathcal{E_S}$ is in off-state, its consumed power is equal to zero. Therefore, consumed power cost of link $(u,v)$ can be modeled as
% \begingroup\makeatletter\def\f@size{8}\check@mathfonts
%     \begin{align*}
%         a(u,v)=
%         \left\{\begin{array}{ll}
%         P_{(u,v)}^{\text{max}}, &\text{\small link }(u,v)\in \mathcal{E}\backslash \mathcal{E_{S}}~\text{\small or link }(u,v)\in \mathcal{E_S}~\text{\small  is in on-state} ,\\
%         0, &\text{\small otherwise.}
%         \end{array}\right.
%     \end{align*}
% \endgroup
\begingroup
\begin{align*}
	a(u,v)=
	\left\{\begin{array}{ll}
		P_{(u,v)}^{\text{max}}, & (u,v)\in \mathcal{E}\backslash \mathcal{E_{S}}~\text{\small or }(u,v)\in \mathcal{E_S}~\text{\small  is in on-state} , \\
		0,                      & \text{\small otherwise.}
	\end{array}\right.
\end{align*}
\endgroup

From the above definitions, the presented system model has included different modes of SDN/NFV networks, e.g., traditional, (partial)SDN, (hybrid)NFV, partial SDN/hybrid NFV, partial SDN/full NFV, and full SDN/full NFV networks.

%%%%%%%%%%%%%%%%%%Problem Formulation %%%%%%%%%%%%%%%%%%%%%%
\section{Problem Formulation}\label{problem}

In this section, we introduce a problem formulation for energy-efficient resource allocation, where the objective is to minimize the number of active nodes and links. To state this formulation mathematically, we initiate our discussion by introducing practical constraints in our proposed model.

When a node $u$ is in off-state, all physical connected links to this node should be in off-states, and vice versa. This constraint is represented as
\begin{equation*}
	\text{C1:}\quad \beta(u,v)\leq \alpha(u), \quad \forall u,v\in \mathcal{N}, \forall (u,v)\in \mathcal{E},
\end{equation*}
\begin{equation*}
	\text{C2:}\quad \alpha(u) \leq \sum_{\{v| (u,v)\in \mathcal{E}\}}\beta(u,v), \quad \forall u\in \mathcal{N},
\end{equation*}
where C1 forces all links  connected to node $u$ to be in off-states if node $u$ is in off-state, and C2 is utilized to turn off node $u$ if all related links are in off-states.

To select node $u$ for running an NF $g_{\bar{k}}^f$ of flow $f$, the node must be able to execute the NF. We can represent this condition as

\begin{equation*}
	\text{C3:}\quad \mu_{u}^{\bar{k}f}\leq\mu_{u}^{k},\quad %\chi^f_{k\rightarrow \bar{k}}(g_k)=g^f_{\bar{k}},
	\forall g_{\bar{k}}^f\in \mathcal{G}^{f}, \forall u\in \mathcal{N_{M}}\cup \mathcal{N_{N}}, \forall f.
\end{equation*}

Here, we assume that NF splitting is not allowed. Therefore, each NF of flow $f$ must be run entirely in one node. Mathematically, this point can be represented as
\begin{equation*}
	\text{C4:}\quad \sum_{u\in \mathcal{N_{M}}\cup\mathcal{N_{N}}} \mu_{u}^{\bar{k}f}=1,\quad \forall g_{\bar{k}}^f\in \mathcal{G}^{f}, \forall f,
\end{equation*}
meaning that each $g_{\bar{k}}^f$ is only run in one node \cite{MAT}.

The sum of required capacities of type $l$ of placed VNFs in node $u$ should not exceed from the capacity of that node. We show this capacity limitation as
\begin{equation*}
	\text{C5:}\quad  \sum_{k=1}^{K} \mu_{u}^{k} c_{k}^{l}\leq \alpha (u)c_{u}^{l}, \quad \forall l,~\forall u\in \mathcal{N_{N}}.
\end{equation*}
where right and left sides of inequality  are the capacity of type $l$  and the used capacity of type $l$ of node $u$, respectively.

Another important point is when  each flow $f$ passes through some of NFs, e.g., tunneling and encryption, data rate of flow $f$ will be changed due to additional overheads and signaling procedures in wireless networks \cite{gsmlte}. Therefore, in this context for each $g^f_{\bar{k}}$, it is assumed that there exists $\gamma^f_{\bar{k}} \geq 0$, such that when the incoming data rate of flow $f$ to  $g^f_{\bar{k}}$ is equal to $r_f$, the outgoing rate of flow $f$ from $g^f_{\bar{k}}$ is equal to $\gamma^f_{\bar{k}}r_f$. Thus, the ingress capacity limitation of non-NFV node $u\in \mathcal{N_{M}}$ is defined as
\begin{equation*}
	\text{C6:}\quad \sum_{\forall f} \sum_{k \in \mathcal{G}^{f}} \zeta_{u}^{\bar{k}f}  \mu_{u}^{\bar{k}f} r^{f} \prod_{j=1}^{\bar{k}-1} \gamma^{f}_j \leq \alpha(u) r_{u},  \quad  \forall u\in \mathcal{N_{M}}.
\end{equation*}
where $r^{f} \prod_{j=1}^{\bar{k}-1} \gamma^{f}_j$ is ingress rate of flow $f$ for the  $k^{th}$ function. $\mu_{u}^{\bar{k}f}$  ensures that the function is executed at node $u$ and $\zeta_{u}^{\bar{k}f}$  guarantees that the previous function is not executed at  node $u$. For nodes $u\in \mathcal{N_{N}}$, there is no need to check that the previous function is executed at those nodes. Therefore,
the ingress capacity limitation of VNF $k$  placed in node $u\in \mathcal{N_{N}}$ can be expressed as

\begin{equation*}
	\text{C7:}\quad \sum_{\forall f}   \mu_{u}^{\bar{k}f} r^{f} \prod_{j=1}^{\bar{k}-1} \gamma^{f}_j \leq r_{k}, \quad  %\chi^f_{k\rightarrow \bar{k}}(g_k)=g^f_{\bar{k}},~
	\forall g_k \in \mathcal{G},~\forall u\in \mathcal{N_{N}}.
\end{equation*}

After determining the places where all $ g_{\bar{k}}^f\in \mathcal{G}^{f}$ in the network are run, one complete path between nodes should be assigned. To set this point, first, the source $s_f$ should be connected to the node $u$ running $g_1^f$. This constraint can be represented as
\begin{equation*}
	\text{C8:}\quad \sum_{j=1}^{J_{s^{f}\rightarrow u}}\Gamma_{s^{f} \rightarrow u}^{fj}=\mu_{u}^{1f}, \quad \forall f, \forall u \in \mathcal{N_{M}}\cup \mathcal{N_{N}}.
\end{equation*}
where $\mu_{u}^{1f}$ denotes whether the first function of flow $f$ is executed in node $u$ or not. If it is executed, from the source to node $u$  only one path is selected.
Then, all other nodes running $ g_{\bar{k}}^f\in \mathcal{G}^{f}$ should be connected based on the order of $ g_{\bar{k}}^f$ in $\mathcal{G}^f$. This consecutive ordering of paths can be represented as

\begin{equation*}
	\text{C9:}\quad  \sum_{j=1}^{J_{u\rightarrow v}}\Gamma_{u \rightarrow v}^{fj}=\mu_{u}^{\bar{k}-1f}.\mu_{v}^{\bar{k}f}, \quad \forall f, \forall u,v \in \mathcal{N_{M}}\cup \mathcal{N_{N}},
\end{equation*}
meaning that only one path is selected between the nodes which run consecutive NFs. Finally, the destination should be connected  to the node running the last NF ($ g_{\bar{k}}^f$) of flow $f$. This constraint is also represented as
\begin{equation*}
	\text{C10:}\quad  \sum_{j=1}^{J_{u\rightarrow d^{f}}}\Gamma_{u \rightarrow d^{f}}^{fj}=\mu_{u}^{\bar{K}f} \quad \forall f, \forall u \in \mathcal{N_{M}}\cup \mathcal{N_{N}}.
\end{equation*}
Consequently, from C8-C10, one complete path between the source and the destination of each flow is created. To provide reliable communication for
different flows in the network, besides limitations on links capacity, we should consider an upper limit on link utilization to prevent unbounded queuing
delay. The capacity used in link $(u,v)$ is calculated  in \eqref{C_def}. Each link $(u,v)$ is located in the path between two nodes $u'$ and $v'$, where they are executed at two successive functions of flow $f$ , respectively. Thus, the ingress rate of node $v'$ is equal to the rate passed through  link $(u,v)$. In \eqref{C_def}, $\zeta_{v'}^{\bar{k}f} r^{f} \prod_{j=1}^{\bar{k}-1} \gamma^{f}_{\bar{k}}$ calculates ingress rate of node $v'$ which is equal to the rate of link $(u,v)$. $\mu_{v'}^{\bar{k}f}$ ensures that the $k^{th}$ function of flow $f$  runs in node $v'$. $\Gamma_{u'\rightarrow v'}^{fj}$  and $\rho_{u \rightarrow v}^{j(u',v')}$ specify that the $j^{th}$ path is selected for $u'\rightarrow v'$ of flow $f$ and   link $(u,v)$  is  in  $j^{th}$ path, respectively. Finally, to calculate the amount of capacity used by link $(u,v)$, all functions and flows are added together.

\begin{figure*}
	\begin{equation}\label{C_def}
		%\begin{align*}
		% \vspace{-0.2in}
		c^c(u,v)=
		% \end{align*}
		%\begin{align*}
		\sum_{\forall f} \sum_{g_{\bar{k}}^f\in \mathcal{G}^{f}} \sum_{v'\in \mathcal{N_{M}}\cup\mathcal{N_{N}},v'\neq u'} \sum_{u'\in \mathcal{N_{M}}\cup\mathcal{N_{N}}} \sum_{j=1}^{J_{u'\rightarrow v'}}
		%\end{align*}
		%\begin{align*}
		\rho_{u'\rightarrow v'}^{j(u,v)} \Gamma_{u'\rightarrow v'}^{fj} \mu_{v'}^{\bar{k}f}\zeta_{v'}^{\bar{k}f} r^{f} \prod_{j=1}^{\bar{k}-1} \gamma^{f}_{\bar{k}}.
		%\end{align*}
	\end{equation}
	\vspace{-0.2in}
\end{figure*}

In order to avoid congestion in the links, we consider a margin for the maximum links capacity and define $0<\tau(u,v)\leq 1$ as a utilization factor of link $(u,v)$. Then, for link $(u,v) \in \mathcal{E}$, we add the following constraint:

\begin{equation*}
	\text{C11:}~ c^c(u,v)\leq \tau(u,v) \beta(u,v) c(u,v), ~ \forall (u,v)\in \mathcal{E}.
\end{equation*}
where $c^c(u,v)$ denotes the used capacity of link $(u,v)$ and must be less than the multiplication of the link capacity $c(u,v)$, the link state $\beta(u,v)$, and the link utilization  $\tau(u,v)$.

To reach energy efficiency in the proposed framework, we define a new utility function  as
\begin{equation*}
	Z(\alpha (u),\beta (u,v),  \mu _{u}^{k},\mu _{u}^{\bar{k}f},\Gamma _{u \to v}^{fj}) =\sum_{u\in \mathcal{N} \backslash\mathcal{N_{T}}} a(u)\alpha(u)+\sum_{(u,v)\in \mathcal{E_{S}}} a(u,v)\beta(u,v),
\end{equation*}

%\begingroup\makeatletter\def\f@size{7}\check@mathfonts
%\begin{align*}
%	Z(\alpha (u),\beta (u,v), & \mu _{u}^{k},\mu _{u}^{\bar{k}f},\Gamma _{u \to v}^{fj}) =                                                        \\
%	                          & \sum_{u\in \mathcal{N} \backslash\mathcal{N_{T}}} a(u)\alpha(u)+\sum_{(u,v)\in \mathcal{E_{S}}} a(u,v)\beta(u,v),
%\end{align*}
%\endgroup
where the first term of the objective function shows the sum of the consumed power of  all nodes except non-SDN nodes, i.e., $u\in \mathcal{N} \backslash \mathcal{N_{T}}$, and, the second term represents the sum of power consumed by all SDN links $(u,v) \in \mathcal{E_{S}}$. Given the above objective function  and the constraints from C1-C11, the problem of energy efficient optimization is formulated in \eqref{Approximation of Power alloction problem}.

\begin{figure*}
	\begin{equation}
		\label{Approximation of Power alloction problem}
		\begin{aligned}
			 & \underset{\alpha (u),\beta (u,v),\mu _{u}^{k},\mu _{u}^{\bar{k}f},\Gamma _{u \to v}^{fj}}{\text{min}}
			 &                                                                                                       & Z(\alpha (u),\beta (u,v),\mu _{u}^{k},\mu _{u}^{\bar{k}f},\Gamma _{u \to v}^{fj})                                                           \\
			 & \quad \quad \text{subject to}
			 &                                                                                                       & \quad \quad \quad \quad \text{C1-C11}                                                                                                       \\
			 &
			 &                                                                                                       & \alpha (u),\beta (u,v),\zeta _{u}^{\bar{k}f},\mu _{u}^{k},\mu _{u}^{\bar{k}f},\rho _{u \to v}^{,(u,v)},\Gamma _{u \to v}^{fj} \in \{ 0,1\}. \\
		\end{aligned}
	\end{equation}
	\hrule
\end{figure*}

All variables of the optimization problem \eqref{Approximation of Power alloction problem} are integers; thus, the problem is an integer programming.
Several variables are multiplied together in C11, and so the optimization problem is non-linear.  Therefore, the optimization problem belongs to the class of non-linear integer programming (NLIP) problems, which are known to be computationally complex. Even for simpler scenarios, for a large set of predefined paths, it is not trivial to solve the problem \cite{enhancingLakshman}. Consequently, proposing efficient algorithms with tractable computational complexity is essential. To reach this goal, we propose the Modified Viterbi Algorithm (MVA) to determine the place where all NFs are run and the routing between the nodes. In our proposed algorithm, instead  of selecting one path for VA in each stage \cite{coding}, we store $\Psi^f_{\bar{k}}$-tuple of paths in stage $\bar{k}$ for some integer $\Psi^f_{\bar{k}}$. We will show that such modification can considerably increase the energy efficiency of this type of network. At the same time, the complexity of the algorithm remains polynomial in time.
In Tables \ref{table1} and \ref{variableTable}, the parameters and the variables are summarized, respectively.

\begin{table*}
\vspace{-0.2in}
\caption{Parameters}
\label{table1}
\vspace{-.1in}
\begin{minipage}[t]{0.5\textwidth}
\begingroup
%\scriptsize
\begin{tabular}[t]{>{\centering\arraybackslash}p{0.15\textwidth}p{0.77\textwidth}}
	\hline

	%\cellcolor{gray} {\bf Symbol} & \cellcolor{gray} {\bf Description}\\
	{\bf Symbol}                                                                  & {\bf Description}                                                      \\

	\hline
	%	\centering
	%	\begin{tabular}{c l|}
	%		\hline
	%		\cellcolor{gray}{\bf Notations} & \cellcolor{gray}{\bf Description}\\ \hline

	$\mathcal{N}$                                                                 & Set of nodes                                                           \\
	$\mathcal{N_{T}}$                                                             & Set of non-SDN switches                                                \\
	$\mathcal{N_{S}}$                                                             & Set of SDN switches                                                    \\
	$\mathcal{N_{M}}$                                                             & Set of non-NFV nodes                                                   \\
	$\mathcal{N_{N}}$                                                             & Set of NFV enabled nodes                                               \\
	$\mathcal{E}$                                                                 & Set of links                                                           \\
	$\mathcal{E_{S}}$                                                             & Set of SDN links                                                       \\
	$\mathcal{G}^{f}$                                                             & Set of NFs                                                             \\
	$\mathcal{G}^{f}$                                                             & Set of NFs of flow $f$                                                 \\
	$\mathcal{G}_{u}$                                                             & Set of NFs of $u\in \mathcal{N_{M}}$                                   \\
	$\mathcal{C}_{u}$                                                             & Set of resources capacity of server node $u$                           \\
	$\mathcal{C}_{k}$                                                             & Set of resources requirements of VNF $k$                               \\
	$\mathcal{P}_{u \rightarrow v}$                                               & Set of all paths between nodes $u$ and $v$                             \\
	$\mathcal{P}^{f*}_{\tilde{u}_{\bar{k}-1}^f\rightarrow \tilde{u}_{\bar{k}}^f}$ & Path related to edge $(\tilde{u}_{\bar{k}-1}^f,\tilde{u}_{\bar{k}}^f)$ \\
	$\Pi_{\bar{k}}^f$                                                             & Set of selected nodes in stage $s_{\bar{k}}^f$ for flow $f$            \\
	$\tilde{C}$                                                                   & Capacity matrix of a candidate path                                    \\
	$\bar{K}^f$                                                                   & Number of NFs for flow $f$                                             \\

	$a(u)$                                                                        & Power cost of  $u$                                                     \\
\end{tabular} \endgroup
\end{minipage} \hfill \vline
%\vrule
\begin{minipage}[t]{0.5\textwidth}
\begingroup
%\scriptsize
\begin{tabular}[t]{>{\centering\arraybackslash}p{0.15\textwidth}p{0.77\textwidth}}

	\hline
	%\cellcolor{gray} {\bf Symbol} & \cellcolor{gray} {\bf Description}\\
	{\bf Symbol}                         & {\bf Description}                                                                \\

	\hline
	$\gamma_{\bar{k}}^f$                 & Data rising factor of NF $g_{\bar{k}}^f$                                         \\
	$c_{u}^{l}$                          & Resource capacity of  type $l$ of node $u$                                       \\
	$c_{k}^{l}$                          & Resources requirements of type $l$  VNF $k$                                      \\
	$r_{u}$                              & Maximum ingress traffic flow of node $u \in \mathcal{N_{N}}\cup \mathcal{N_{M}}$ \\
	$r_{k}$                              & Maximum ingress traffic flow of VNF $k$                                          \\
	$P_{u}^{\text{max}}$                 & Maximum consumed power of node $u $                                              \\
	$a(u,v)$                             & Power cost for  $(u,v)$                                                          \\

	$c(u,v)$                             & Capacity of $(u,v)$                                                              \\
	$\tau$                               & Max link utilization                                                             \\
	$r^{f}$                              & Traffic of flow $f$                                                              \\
	$s^{f}$                              & Source of flow $f$                                                               \\
	$d^{f}$                              & Destination of flow $f$                                                          \\
	$s^f_{\bar{k}}$                      & Stage $\bar{k}$ of flow $f$                                                      \\
	$g_{\bar{k}}^{f}$                    & $k^{th}$ NF of flow f                                                            \\

	$\rho^{j,(u',v')}_{u \rightarrow v}$ & To show if  $(u',v')$ exists in $\mathcal{P}^{j}_{u \rightarrow v} $             \\
	$w^f(u)$                             & Weight of $u$                                                                    \\
	$w^f(u,v)$                           & Weight of $(u,v)$                                                                \\
\end{tabular}
\endgroup

\end{minipage}
\hrule width 1.006\textwidth
%\vspace{-0.3in}

\end{table*}

\begin{table}

	\caption{Variables}
	\label{variableTable}
	\vspace{-.1in}
	\centering
	\begin{tabular}{cl}
		\hline
		Variable                         & Description                                                                                                 \\
		\hline
		$\Psi^f_{\bar{k}}$               & Number of stored paths in stage $s^f_{\bar{k}}$ for $u$                                                     \\
		$\alpha(u)$                      & Power state of $u$                                                                                          \\
		$\beta(u,v)$                     & Power state for $(u,v)$                                                                                     \\
		$\mu_{u}^{k}$                    & To choose for node $u\in \mathcal{N_{N}} $ running VNF $k$                                                  \\
		$\mu_{u}^{\bar{k}f}$             & To choose node $u\in \mathcal{N_{N}} $ running NF $g_{\bar{k}}^f$                                           \\
		$\Gamma^{f,j}_{u \rightarrow v}$ & To select path $\mathcal{P}^{j}_{u \rightarrow v}$ for nodes $u$, $v$ from $\mathcal{P}_{u \rightarrow v} $ \\
		\hline
	\end{tabular}
	\vspace{-.1in}

\end{table}

%%%%%%%%%%%%%%%%%%%%%%%%%%%%%%%%%The Proposed Algorithm%%%%%%%%%%%%
\section{The Proposed Algorithm} \label{algorithmsection}%\vspace{-0.5in}

In this section, we introduce the steps of the proposed algorithm to solve the optimization problem. First, we present the weight assignment strategy based on the on and off-states and types of the nodes and the links. The weights are determined based on the energy consumption of the nodes or the links. If a node or a link is used, its weight is assigned by the amount of increased energy consumption of the network.
The network is modeled by a multi-stage graph for each flow. The sets of candidate nodes for running NFs in each stage and  paths between the stages are determined based on MVA.
This algorithm is briefly demonstrated in Fig. \ref{Example1} and Fig. \ref{flowchart} plots the flowchart of the algorithm.  The algorithm is run for each flow. We assume that for the first flow, i.e., $f=1$, all nodes and links are in off-states. The weight assignment strategy of all nodes and links is updated for each flow.

\subsection{Weight Assignment Strategy} \label{weighting}
We consider the weight assignment of each node $u\in \mathcal{N_{S}} \cup \mathcal{N_{T}}$ for flow $f$ as
\begin{equation*}
	w^f(u)=
	\left\{\begin{array}{ll}
		\varepsilon(u),   & $if $ u\in \mathcal{N_{T}} ~$or$~ u\in \mathcal{N_S}$ is in on-state,$ \\
		P_u^{\text{max}}, & $if $ u\in \mathcal{N_S}~$ is in off-state,$
	\end{array}\right.
\end{equation*}
where $0<\varepsilon(u)\ll 1,$ and $ ~\varepsilon(u)<a(u)$.
Using this weight assignment strategy, we try to force the algorithm to utilize the resources of nodes $u\in\mathcal{N_S}\cup \mathcal{N_{T}}$ which are in on-state. This is because for switches in our setup, we assume that the consumed power depends on the states (on/off) of a switch and it is not related to the load of the switch. Consequently, this  weight assignment strategy tries to keep off-state nodes in the same states.

For the nodes running NFs, in addition to the difference between the power consumed in off-state and on-state, power consumption in on-state also depends on traffic load of that node. Accordingly, we define the following weight assignment strategy
\begingroup
\begin{align*}
	w^f(u)=
	\left\{\begin{array}{ll}
		\theta P_u^{\text{max}}  ,                   & \text{\small if }  u\in \mathcal{N_{N}}\cup \mathcal{N_{M}} ~\text{\small is in off-state,}  \\
		(1-\theta)\frac{r^f}{r_{u}}P_u^{\text{max}}, & \text{\small if }  u\in \mathcal{N_{N}}\cup \mathcal{N_{M}}  ~\text{\small is  in on-state,}
	\end{array}\right.
\end{align*}
\endgroup
From practical considerations, we have $\theta P_u^{\text{max}}\gg(1-\theta)\frac{r^f}{r_{u}}P_u^{\text{max}}$. Therefore, still via this strategy, we insist to keep the off-state nodes in those states.

Similarly, for each link $(u,v)$, we consider the following weight assignment strategy
\begingroup
\begin{align*}
	w^f(u,v)=
	\left\{\begin{array}{ll}
		\varepsilon(u,v) , & \text{\small if }(u,v)\in \mathcal{E}\backslash\mathcal{E_S}~\text{\small or  }(u,v)\in \mathcal{E_{S}} ~\text{\small is in on-state,} \\
		a(u,v)  ,          & \text{\small if }(u,v)\in \mathcal{E_{S}} ~\text{\small is  in  off-state,}
	\end{array}\right.
\end{align*}
\endgroup
where $0<\varepsilon(u,v)\ll 1, ~\varepsilon(u,v)\ll a(u,v)$. Remarkably, since orchestrator cannot control the on-state and off-state of $u\in \mathcal{N_{T}}$ and $(u,v)\in \mathcal{E}\backslash\mathcal{E_S}$, and since the configuration of these nodes and links from off-state to on-state is time consuming, it is preferable to keep these nodes in on-states. Therefore, in our weight assignment strategies, the weights of these nodes and links are assigned to $w^f(u)=\varepsilon(u)$, $w^f(u,v)=\varepsilon(u,v)$, respectively.

\subsection{Proposed Algorithm Based on Multi-Stage Graph Modeling}

\begin{figure}
	\begin{center}
		\includegraphics[width=0.8\textwidth]{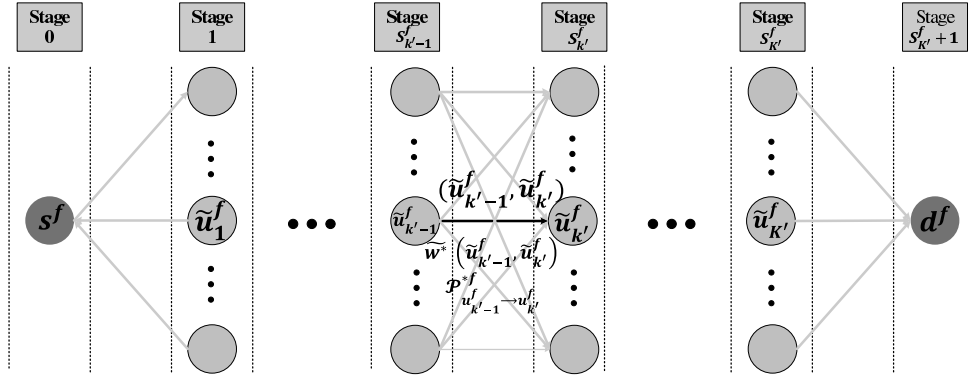}
	\end{center}
	\renewcommand{\captionfont}{\small}
	\caption{Multi-stage graph modeling: for each nodes in two consecutive stages, MVA find edge $(\tilde{u}_{\bar{k}-1}^f,\tilde{u}_{\bar{k}}^f)$ with weight $\tilde{w}^{f*}_{\bar{k}}(\tilde{u}_{\bar{k}-1}^f,\tilde{u}_{\bar{k}}^f)$ and $\mathcal{P}^{f*}_{\tilde{u}_{\bar{k}-1}^f\rightarrow \tilde{u}_{\bar{k}}^f}$.}
	\label{Example1}
	\vspace{-0.2in}
\end{figure}

\begin{figure}
	\includegraphics[width=1\textwidth]{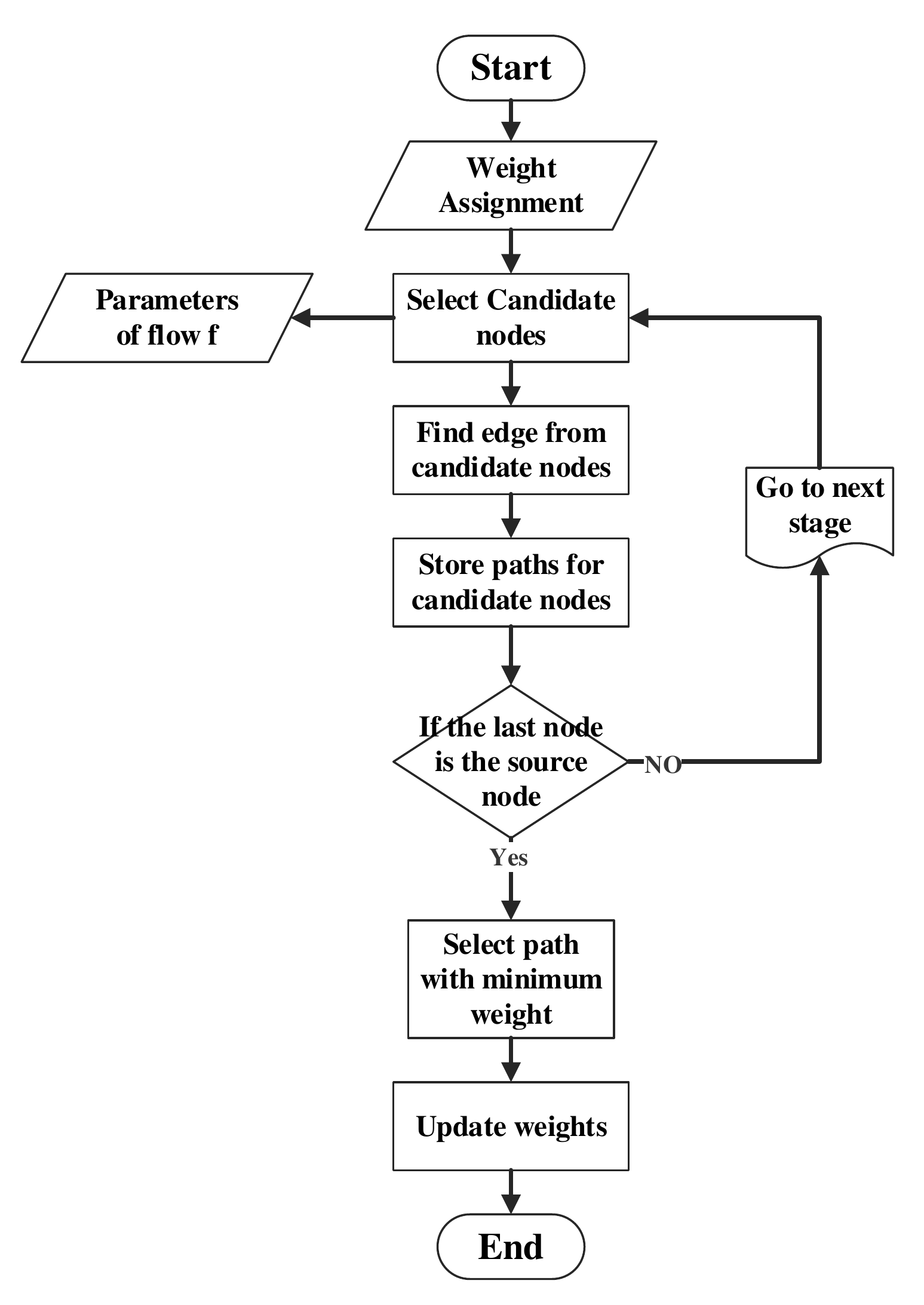}
	\renewcommand{\captionfont}{\small}
	\vspace{-0.2in}
	\caption{Flowchart of the algorithm.}
	\label{flowchart}
	\vspace{-0.2in}
\end{figure}

After the weight assignment strategy, for each flow $f$, a set of nodes running a set $\mathcal{G}^f$ should be determined. Then, the NFs and their related nodes should be connected through network paths based on the order of NFs in $\mathcal{G}^f$, which are determined by the orchestrator. The placement of NFs and routing between nodes are two major interrelated tasks for our algorithm. To jointly perform these two tasks, we resort to the multi-stage graph model \cite{boutaba}. In our algorithm, a set of candidate nodes enabled to run $g^f_{\bar{k}}$ is the candidate set of stage $\bar{k}^f$, as shown in Fig. \ref{Example1}.  Let the  stage of $\bar{k}_f$ to be denoted as $s_{\bar{k}}^f$. We consider source and destination as two predetermined stage of $s^f_0$ and $s^f_{\bar{K}+1}$, respectively. Therefore, a multi-stage graph in our algorithm has $\bar{K}_f+2$ stages. Between two consecutive stages, corresponding to two consecutive NFs in $\mathcal{G}^f$, i.e., $g^f_{\bar{k}}$ and $g^f_{\bar{k}+1}$, there exists a set of edges related to paths between candidate nodes of stages  $\bar{k}_f$ and $\bar{k}_f+1$.

Note that this algorithm is executed for each flow $f$, and, between each node in stages $\bar{k}_f$ and each node in stage $\bar{k}_f+1$, only one edge is selected. This edge corresponds to the shortest path based on the weight assignment strategy from Dijkstra's algorithm \cite{dijkstra1959note}, discussed in Section \ref{Routing}.

\subsubsection{Candidate Set Selection}
For each $g^f_{\bar{k}}\in\mathcal{G}^f$, the candidate set should be chosen based on  the capability of nodes to run $g^f_{\bar{k}}$ and their capacities.  In the initial phase of our proposed algorithm, for each flow $f$, we should update the capacity of nodes and links based on the place where running NFs and routing of flows $1$ to $f-1$. In the followings formulas, we show how this update can be performed from C5, C6, and C7. First to verify  which node $u\in \mathcal{N_{N}}$ is capable to run $g^f_{\bar{k}}$ in terms of available capacity, we consider
\begin{equation*}
	\tilde{c}_{u}^{lf}=  \alpha (u)c_{u}^{l}-\sum_{k=1}^{K} \mu_{u}^{k} c_{k}^{l},\quad
	\forall l,~\forall u\in \mathcal{N_{N}},
\end{equation*}
which shows the updated capacity limit of node $u\in \mathcal{N_{N}}$ for type $l$ resource of flow $f$.
Therefore, to verify if node $u\in \mathcal{N_{N}}$ can run NF $g^f_{\bar{k}}$, the following modified constraint of C5 can be applied
\begin{equation*}
	\tilde{\text{C}}^f\text{5}:%\mu_{u}^{\bar{k}f}
	c_{k}^{l}\leq \tilde{c}_{u}^{,f} \quad %\chi^f_{k\rightarrow \bar{k}}(g_k)=g^f_{\bar{k}},~
	\forall l,~\forall u\in \mathcal{N_{N}}.
\end{equation*}

Similarly, the ingress capacity of node $u\in \mathcal{N_{M}}$ for flow $f$ is updated according to

\begin{equation*}
	\tilde{r}_{u}^f= \alpha(u)r_u-\sum_{ f=1}^{f-1} \sum_{k \in \mathcal{G}^{f}} \zeta_{u}^{\bar{k}f}  \mu_{u}^{\bar{k}f} r^{f} \prod_{j=1}^{\bar{k}-1} \gamma^{f}_j , ~~\forall u\in \mathcal{N_{M}}.
\end{equation*}

Therefore, to verify if node $u\in \mathcal{N_{M}}$ has appropriate ingress capacity for flow $f$, following modified version of C6 is applied

\begin{equation*}
	\tilde{\text{C}}^f\text{6} :\sum_{k \in \mathcal{G}^{f}} \zeta_{u}^{\bar{k}f}  \mu_{u}^{\bar{k}f} r^{f} \prod_{j=1}^{\bar{k}-1} \gamma^{f}_j \leq  \tilde{r}_{u}^f,  \quad  \forall u\in \mathcal{N_{M}}.
\end{equation*}

Also, the updated ingress capacity  of VNF $k$  placed in node $u\in \mathcal{N_{N}}$ can be expressed as
\begin{equation*}
	\tilde{r}_{\bar{k}u}^f=r_k- \sum_{f=1}^{f-1}   \mu_{u}^{\bar{k}f} r^{f} \prod_{j=1}^{\bar{k}-1} \gamma^{f}_j ,\quad \chi^f_{k\rightarrow \bar{k}}(g_k)=g^f_{\bar{k}},
\end{equation*}
and, the modified ingress capacity limitation of VNF $k$ placed in node $u$ based on C7 can be examined by

\begin{equation*}
	\tilde{\text{C}}^f\text{7}: \sum_{k \in \mathcal{G}^{f}} \zeta_{u}^{\bar{k}f}  \mu_{u}^{\bar{k}f} r^{f} \prod_{j=1}^{\bar{k}-1} \gamma^{f}_j \leq  \tilde{r}_{\bar{k}u}^f,  \quad  \forall u\in \mathcal{N_{M}}.
\end{equation*}

In our algorithm, for each flow $f$, we will verify $\tilde{\text{C}}^f\text{5}$-$\tilde{\text{C}}^f\text{7}$ to derive the candidate set of stage $\bar{k}_f$ which is denoted by $\Pi_{\bar{k}}^f$.

\subsubsection{Path Selection}
After selecting nodes for all stages, the paths between nodes in any two consecutive stages should be determined in order to build the end to end connection between source and destination of flow $f$. First, in order to find these paths, we must update the capacity limitation of all links $(u,v)$ in the network based on the traffic flow of $f=1,\cdots,f-1$,
which is  defined as
\begingroup\makeatletter\def\f@size{11}\check@mathfonts
\begin{equation*}
	\tilde{c}^f(u,v)=\tau(u,v) \beta(u,v) c(u,v)-\sum_{f=1}^{f-1} \sum_{g_{\bar{k}}^f\in \mathcal{G}^{f}} \sum_{v'\in \mathcal{N_{M}}\cup\mathcal{N_{N}},v'\neq u'} \sum_{u'\in \mathcal{N_{M}}\cup\mathcal{N_{N}}} \sum_{j=1}^{J_{u'\rightarrow v'}} \rho_{u'\rightarrow v'}^{j(u,v)} \Gamma_{u'\rightarrow v'}^{fj} \mu_{v'}^{\bar{k}f}\zeta_{v'}^{\bar{k}f} r^{f} \prod_{j=1}^{\bar{k}-1} \gamma^{f}_{\bar{k}},
\end{equation*}
\endgroup

where $\tilde{c}^f(u,v)$ is the utilized capacity of link $(u,v)$ by flows $f=1,\cdots,f-1$. Therefore, C11 can be modified as
\begin{equation*}
	\tilde{\text{C}}^f\text{11}: \sum_{g_{\bar{k}}^f\in \mathcal{G}^{f}} \sum_{v'\in \mathcal{N_{M}}\cup\mathcal{N_{N}},v'\neq u'} \sum_{u'\in \mathcal{N_{M}}\cup\mathcal{N_{N}}} \sum_{j=1}^{J_{u'\rightarrow v'}} \rho_{u'\rightarrow v'}^{j(u,v)} \Gamma_{u'\rightarrow v'}^{fj} \mu_{v'}^{\bar{k}f}\zeta_{v'}^{\bar{k}f} r^{f} \prod_{j=1}^{\bar{k}-1} \gamma^{f}_{\bar{k}}\leq \tilde{c}^f(u,v), ~ \forall (u,v)\in \mathcal{E}.
\end{equation*}

Our objective is to minimize the energy consumption of the network. So, next, we need to propose an algorithm to find a path with the least weight, based on the energy consumption of nodes and paths. To reach this goal, we develop a Modified Viterbi Algorithm (MVA) in which an MDRA is applied to find only one path between nodes of each two consecutive stages. Our proposed MDRA will be explained in Section \ref{Routing}.

For two consecutive stages $s^f_{\bar{k}-1}$ and $s^f_{\bar{k}}$, we obtain edge $(\tilde{u}_{\bar{k}-1}^f,\tilde{u}_{\bar{k}}^f)$ for $\tilde{u}_{\bar{k}-1}^f\in \Pi_{\bar{k}-1}^f$, $\tilde{u}_{\bar{k}}^f\in \Pi_{\bar{k}}^f$ based on our proposed algorithm in Section \ref{Routing}, where its  output is $\tilde{w}^f_{\bar{k}}(\tilde{u}_{\bar{k}-1}^f,\tilde{u}_{\bar{k}}^f)$ defined as a weight of a path $\mathcal{P}^f_{\tilde{u}_{\bar{k}-1}^f\rightarrow \tilde{u}_{\bar{k}}^f}$ for edge $(\tilde{u}_{\bar{k}-1}^f,\tilde{u}_{\bar{k}}^f)$.

In stage $s^f_{\bar{k}}$, MVA intends to choose a number of $\psi^f_{\bar{k}}$ paths from $s^d$ to $\tilde{u}_{\bar{k}}^f$ with the smallest weights, $\forall \tilde{u}_{\bar{k}}^f\in \Pi_{\bar{k}}^f$. %}
To do so, for each node $\tilde{u}_{\bar{k}-1}^f\in \Pi_{\bar{k}-1}^f$ and each node $\tilde{u}_{\bar{k}}^f\in \Pi_{\bar{k}}^f$, MVA considers the edge $(\tilde{u}_{\bar{k}-1}^f,\tilde{u}_{\bar{k}}^f)$. Consequently, the candidate paths leading to $\tilde{u}_{\bar{k}}^f$ are the union of the ones stored in each of nodes $\tilde{u}_{\bar{k}-1}^f\in \Pi_{\bar{k}-1}^f$ extended by $\mathcal{P}^f_{\tilde{u}_{\bar{k}-1}^f\rightarrow \tilde{u}_{\bar{k}}^f}$, with their corresponding weights computed as the summation of the weights of the two distinct parts of the path, i.e., $\tilde{w}^f_{\bar{k}}(\tilde{u}_{\bar{k}-1}^f,\tilde{u}_{\bar{k}}^f)$ added to the weight of the path stored in $\tilde{u}_{\bar{k}-1}^f$. Then, the paths eligible to be stored in node $\tilde{u}_{\bar{k}}^f$ are the $\psi^f_{\bar{k}}$ paths with the smallest weights  chosen from the aforementioned paths.

\begin{algorithm}
	
	\begingroup
	\small

\caption{Modified Viterbi Algorithm (MVA)}

\label{Alg1}
%\vspace{-0.04in}
\textbf{Input:} $\mathcal{A}$, $\mathcal{G}$, $\mathcal{G}_u$, $\mathcal{C}_{g_k}$, $\mathcal{C}_u$, $f=(s^{f},d^{f},r^{f},\mathcal{G}^{f})$, $w^f(u)$, $ ~w^f(u,v),~ c(u,v),~ \tilde{c}^f(u,v),~\beta(u,v)$, $a(u,v)$ $~\forall (u,v)\in \mathcal{E}$,  $\alpha(u),~a(u),~\mu_{u}^k,~,\mu_{u}^{\bar{k}f},$ $r_u,~\tilde{r}_u^{f},$ $r_k,~\tilde{r}^f_{\bar{k}u},$ $\tilde{c}_u^{lf},\forall u\in \mathcal{N}$, $~\forall ~k$\\
\textbf{Output:} A path from $s^{f}$ to $d^{f}$, $\alpha (u)$, $\beta (u,v)$, $\mu _{u}^{k}$, $\mu _{u}^{\bar{k},f}$ and $\Gamma _{u \to v}^{f,j}$\\
%\KwResult{Write here the result }
%initialization\;
\For{$s^f_{\bar{k}}$=$1$:$\bar{K}$}
{\For{$\forall u\in \mathcal{N_{M}}$}
{\If{($\tilde{\text{C}}^f6$ holds) \& ($g_{\bar{k}}^{f}\in \mathcal{G}_{u}$)}
{Add $u$ to $\Pi_{\bar{k}}^f$\;
Find edge $(\tilde{u}_{\bar{k}-1}^f,u)$ and its weight with MDRA $\forall \tilde{u}_{\bar{k}-1}^f\in \Pi_{\bar{k}-1}^f $\;
Store $\psi^f_{\bar{k}}$ paths with least weights from a set of candidate paths\;
}
}

\For{$\forall u\in \mathcal{N_{N}}$}
{\If{($\mu_{u}^{\bar{k}f}=1$) \& ($\tilde{\text{C}}^f7$ holds)}
{Add $u$ to $\Pi_{\bar{k}}^f$\;
Find edge $(\tilde{u}_{\bar{k}-1}^f,u)$ and its weight with MDRA $\forall \tilde{u}_{\bar{k}-1}^f\in\Pi_{\bar{k}-1}^f $\;
Store $\psi^f_{\bar{k}}$ paths with least weights from a set of candidate paths\;
}
\If{($\mu_{u}^{\bar{k}f}=0$) \& ($\tilde{\text{C}}^f5$ holds)}
{
Add $u$ to $\Pi_{\bar{k}}^f$\;
Find edge $(\tilde{u}_{\bar{k}-1}^f,u)$ and its weight with MDRA $\forall \tilde{u}_{\bar{k}-1}^f\in \Pi_{\bar{k}-1}^f $\;
\For{All candidate paths}{
	\For{l=1:L}{
		\If{$\tilde{\tilde{\text{C}}}^f5$ does not hold}{Remove this path from set of candidate paths\;}
	}
}
Store $\psi^f_{\bar{k}}$ paths with least weights from candidate paths\;
Update $\tilde{C}$ of stored paths\;
}
}

}
Find edge $(\tilde{u}_{\bar{k}}^f,d_f)$ with MDRA $\forall \tilde{u}_{\bar{k}-1}^f\in\Pi_{\bar{k}}^f $\;
Find weights of candidate paths to $d_f$\;
Select a path with smallest weight\;
\textbf{Return} $\mathcal{P}^f$ and variables $\alpha (u)$, $\beta (u,v)$, $\mu _{u}^{k}$, $\mu _{u}^{\bar{k},f}$ and $\Gamma _{u \to v}^{f,j}$\;
Update all weights and $\tilde{\text{C}}^f5-\tilde{\text{C}}^f7$\;
\endgroup

\end{algorithm}

As mentioned before there is a chance to select a specific node $u \in\mathcal{N_{N}} $ more than once to place more than two VNFs for flow $f$. In this case, we need to update the capacity limitation of each specific node $u\in \mathcal{N_{N}}$ for each selected path of flow $f$ based on the above explained procedure. To address this issue in our algorithm, we define a matrix $\tilde{C}=[\tilde{\tilde{c}}_{u}^{lf}]_{N_N\times L }$, where $\tilde{\tilde{c}}_{u}^{lf}$ is the updated capacity limit of the resource type $l$ of node $u\in \mathcal{N_{N}}$ for flow $f$. This matrix, along with the path's corresponding weight, is stored in node $\tilde{u}_{\bar{k}}^f$. At stage zero, the entries of the matrix are initialized as $\tilde{\tilde{c}}_{u}^{lf}=\tilde{c}_{u}^{lf}$. To meet the constraint on the capacity limit of node $\tilde{u}_{\bar{k}}^f$, in stage $s^f_{\bar{k}}$, the paths satisfying the following constraint remain in the set of candidate paths of node $\tilde{u}_{\bar{k}}^f$, and are evicted otherwise:

\begin{equation*}
	\tilde{\tilde{\text{C5}}}:~~~~ c_{k}^{l}\leq \tilde{\tilde{c}}_{u}^{lf}, \quad
	\forall l.
\end{equation*}
In addition, when a path meets the above constraint, $\tilde{\tilde{c}}_{u}^{lf}$ should be updated as
\begin{equation*}
	\tilde{\tilde{c}}_{u}^{lf}=\tilde{\tilde{c}}_{u}^{lf}-c_{k}^{l}, \quad
	\forall l.
\end{equation*}
For stage $s^f_{\bar{K}+1}$, we have $\psi^f_{\bar{K}+1}$ paths from source to destination. The  path with minimum weight, corresponding to the least energy consumption is selected. The above MVA is summarized in Algorithm \ref{Alg1}.

Finally, the output factors of Algorithm \ref{Alg1} are the selected path between source and destination $\mathcal{P}^f$ as well as variables $\alpha (u)$, $\beta (u,v)$, $\mu _{u}^{k}$, $\mu _{u}^{\bar{k},f}$ and $\Gamma _{u \to v}^{f,j}$.  $w^f(u)$, $w^f(u,v)$, $a(u)$, $a(u,v)$. Afterwards, all capacities are updated according to values of Algorithm \ref{Alg1}.

\subsection{Modified Dijkstra's Routing Algorithm }\label{Routing}
In this section, we explain our proposed MDRA to find an edge between $\tilde{u}_{\bar{k}-1}^f$ and $\tilde{u}_{\bar{k}}^f$ of two consecutive stages $s^f_{\bar{k}-1}$ and $s^f_{\bar{k}}$. Here,  we have a set of the paths $\mathcal{P}_{\tilde{u}_{\bar{k}-1}^f \rightarrow \tilde{u}_{\bar{k}}^f}$, and we want to choose a path with minimum weight corresponding to minimum power consumption  of nodes and links. We resort to Dijkstra's algorithm \cite{dijkstra1959note}. However to consider the capacity limitation of links and the weights of nodes, we propose a modified Dijkstra's algorithm to find a path in graph $\mathcal{A}(\mathcal{N},\mathcal{E})$ for nodes $\tilde{u}_{\bar{k}-1}^f$ and $\tilde{u}_{\bar{k}}^f$. In our proposed MDRA, we update weight of each link as
\begin{equation*}
	\tilde{w}^f_{\bar{k}}(u,v)=
	\left\{\begin{array}{cl}
		w^f(u,v)+w^f(u)+w^f(v) , & $if$~~ \tilde{\text{C}}^f11 $ holds$, \\
		\infty,                  & $Otherwise$,
	\end{array}\right.
\end{equation*}
where if there is no capacity, the weight of that link is set to infinity, meaning that this link will be avoided to choose for flow $f$\footnote{In Dijkstra's algorithm,  the links only have weights and weights of nodes does not considered. In addition, in Dijkstra's algorithm, the capacity of nodes are not considered while in our proposed algorithm, we consider the capacity of nodes as well  \cite{dijkstra1959note}.}.
In addition, the nodes' weights are considered in routing via summation of  nodes' weights to their corresponding weights of physical  connected links.
In this paper, similar to \cite{MAT,reliabilityfordijkstra}, we develop Dijkstra's algorithm. Consequently, MDRA chooses the shortest path based on Dijkstra's algorithm denoted by $\mathcal{P}^{f*}_{\tilde{u}_{\bar{k}-1}^f\rightarrow \tilde{u}_{\bar{k}}^f}$. Another output of MDRA is  $\tilde{w}^{f*}_{\bar{k}}(\tilde{u}_{\bar{k}-1}^f,\tilde{u}_{\bar{k}}^f)$ corresponding to the weight of $\mathcal{P}^{f*}_{\tilde{u}_{\bar{k}-1}^f\rightarrow \tilde{u}_{\bar{k}}^f}$.

\begin{figure}
	\vspace{-.2in}
	\begin{subfigure}[b]{0.5\textwidth}
		\includegraphics[width=\textwidth]{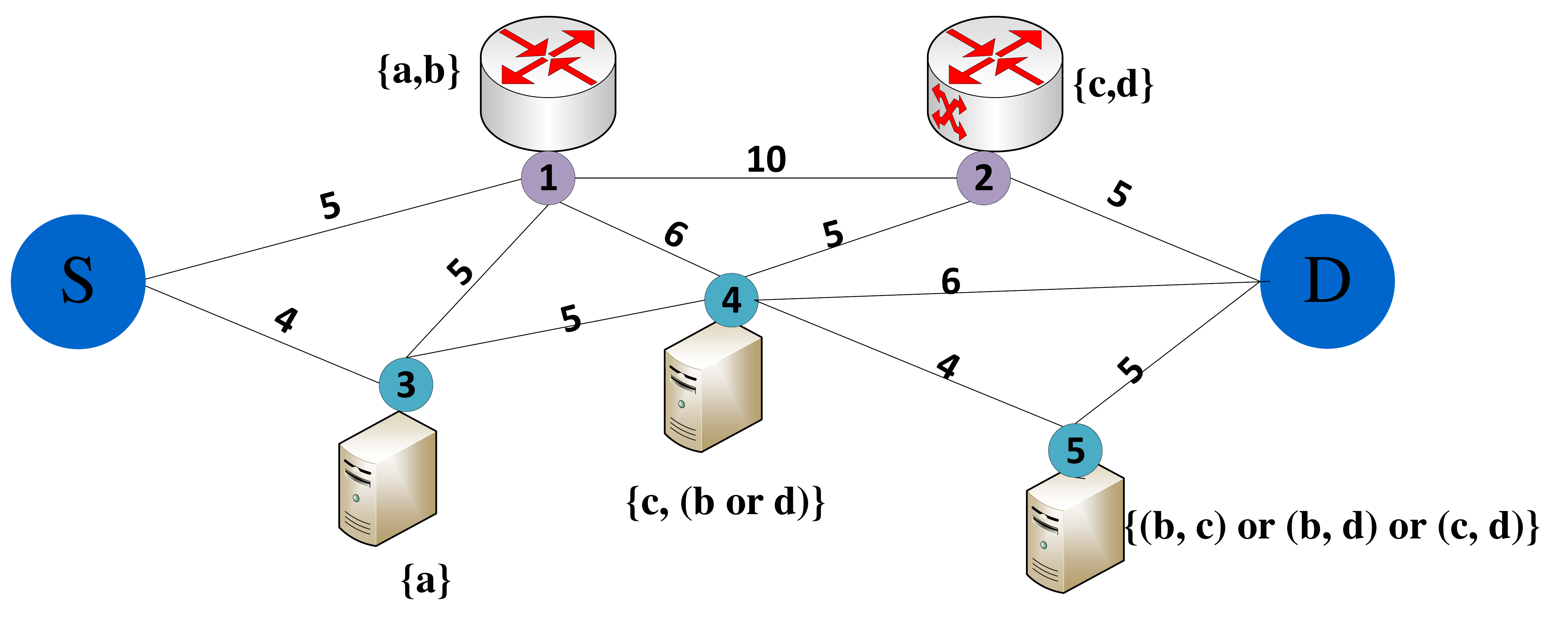}
		\vspace{-.2in}
		\renewcommand{\captionfont}{\small}
		\vspace{-.2in}
		\caption{The topology of the network}
	\end{subfigure}
	\begin{subfigure}[b]{0.5\textwidth}
		\includegraphics[width=\textwidth]{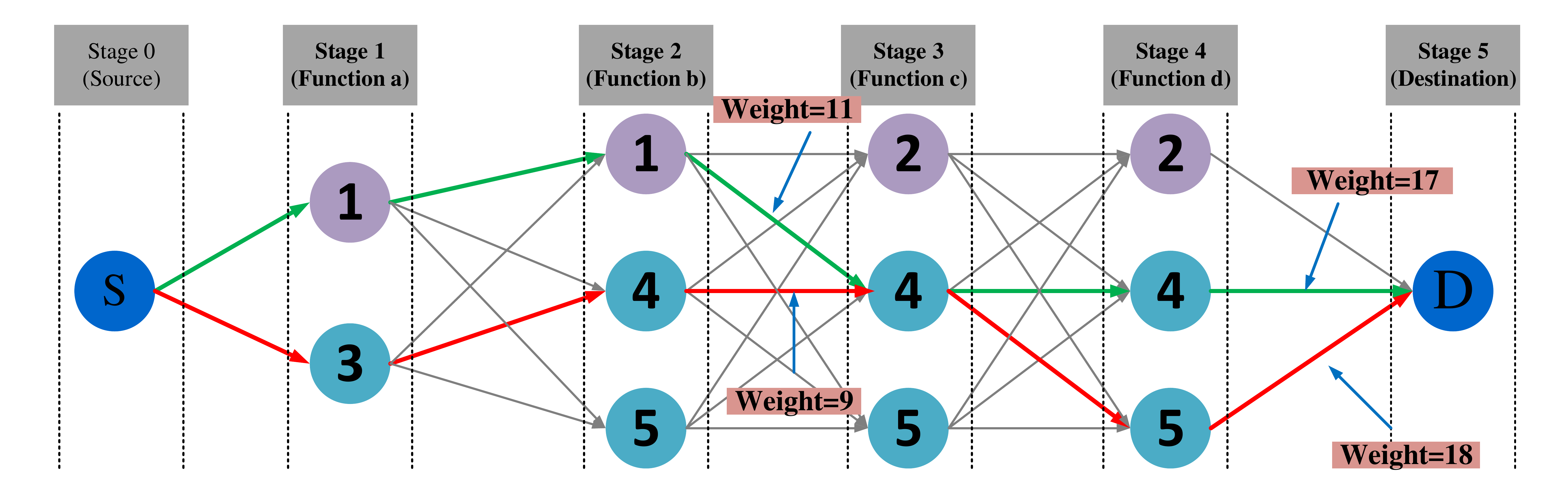}
		\renewcommand{\captionfont}{\small}
		\caption{The stages, the candidate nodes and the routes}
	\end{subfigure}
	\renewcommand{\captionfont}{\small}
	\vspace{-.4in}
	\caption{An illustrative example of proposed algorithm }
	\label{Example2}
	\vspace{-0.4in}
\end{figure}

\subsection{Illustrative Example}
In this section, we present an example of the proposed algorithm for the network of Fig. \ref{Example2}, and also show why we should store more than one path for the nodes. As it can be seen in  Fig. \ref{Example2},  nodes 1 and 2 are non-NFV which run NFs of $\{a,b\}$ and $\{c,d\}$, respectively. Node 3 runs NF $\{a\}$ and cannot run other NFs. NF $\{c\}$  is run on  node 4 and it can be placed one of two NFs $\{b\}$ and $\{d\}$. Node 5 can execute two NFs out of $\{b\}$, $\{c\}$, and $\{d\}$. In this network, the weights of the paths are as shown in Fig. \ref{Example2}.

Consider a flow $f=\{$S,D,100 Mbps,$\mathcal{G}^{f}\}$, where "S" and "D"  denote  the source and the destination of flow $f$, respectively, and $\mathcal{G}^{f}=\{a,b,c,d\}$. At each stage of the proposed algorithm, the nodes which can execute that NF are selected as noted in  Fig. \ref{Example2} (a). For example, in stage one, nodes 1 and 3 are selected, and in stage two, nodes 1, 4, and 3 are chosen. Let us assume that for each stage up to 3 paths can be stored.

For each candidate node in stage $s^f_{\bar{k}}$, its edges to all other nodes in the previous stage are found. Note that for each edge, there exists one path, including one physical link or several connected physical links. For instance, for edge $(1,4)$ between the second and third stages, its corresponding path $1\rightarrow 4$ includes one physical link.  Likewise, for edge $(1,5)$ again between the second and third stages, its corresponding path is $1\rightarrow 4 \rightarrow 5$ including two physical links $(1,4)$ and $(4,5)$.
After verifying $\tilde{\tilde{\text{C5}}}$, $\psi^f_{\bar{k}}\leq 3$, ${\bar{k}}=1,2,3$, the paths with the lowest weights from the source to that node are stored. For example, for node 4 in the third stage, path $S \rightarrow 3 \rightarrow 4 \rightarrow 4$ with weight 9, path $S \rightarrow 1 \rightarrow 1 \rightarrow 4$ with weight 11, and path $S \rightarrow 1 \rightarrow 4 \rightarrow 4$ with weight 11 are stored.

Consider the case that $\psi^f_{\bar{k}}=1$, ${\bar{k}}=1,2,3$, then the red path is selected for candidate node 4 in the third stage. Consequently, in the fourth stage, again, the red path is stored to run function $d$. Finally, the red path from the source to the destination with weight 18 is selected for the flow $f$. However, when $\psi^f_{\bar{k}}\leq 3$, ${\bar{k}}=1,2,3$, the red and the green paths are stored for candidate node 4 in the fourth stage. Finally, the green path from the source to the destination with weight 17 is selected, which has a lower weight than the red path. This illustrative example shows how with MVA and considering three candidate paths instead of one path in VA, the performance of the network can be improved in terms of energy efficiency.

\subsection{Computational Complexity}

In this section, we investigate the computational complexity of the proposed algorithms and compare them with the exhaustive search.
At each stage of the proposed algorithm, $N_M+N_N$ nodes are verified, and for each node,  $O(N_M+N_N)$ edges are found. For each candidate node in each stage, all links (the number of links is $E$) are first verified, and then,  Dijkstra's algorithm runs with complexity $O(E+N\log(N))$ \cite{complexityDijkstra}. Thus, MDRA performs $O(E+N\log(N)+E)=O(2E+N\log(N))$ computations. The number of stages in the algorithm is $|\mathcal{G}^f|$. Therefore, the computational complexity of Algorithm is $O(|\mathcal{G}^f|(N_M+N_N)^2(2E+N\log(N)))$, which is polynomial time.

However, the computational complexity of the exhaustive search algorithm for optimization problem \eqref{Approximation of Power alloction problem} is  $O(e (N-2)! \times 2^{(N_S+N_N)} \times 2^E \times 2^{|\mathcal{G}^f|(N_M+N_N)} )$, where $e$ is Neper number. The first term, i.e., $e (N-2)!$, approximates the entire paths which should be searched for the fully connected graph between two specific nodes. The second term, i.e.,   $2^{(N_S+N_N)}$, is the number of all on/off states of SDN and NFV nodes that should be searched via exhaustive search.  $2^E$ denotes the number of the entire on/off states of SDN links. Finally, to find out which node runs a specific NF, exhaustive search algorithm should search among $2^{|\mathcal{G}^f|(N_M+N_N)}$ possible scenarios among nodes. Comparing the exponential order of the exhaustive search with polynomial order of Algorithm 1 demonstrates its efficiency in terms of computational complexity.

%%%%%%%%%%%%%%%%%%Simulations%%%%%%%%%%%%%%%%%%%%%%%%%%%%%%%%%%%%%%
\section{Simulations}\label{simulation}

In this section, we evaluate the performance of MVA versus different network parameters and scenarios.
We first compare the performance of MVA with the optimal solution in a small network with normalized parameters. Then, we evaluate the performance of MVA in three different sizes of MCNs.

\begin{figure*}
	\vspace{-0.3in}
	\centering
	\begin{subfigure}[b]{0.32\textwidth}
		\includegraphics[width=\textwidth]{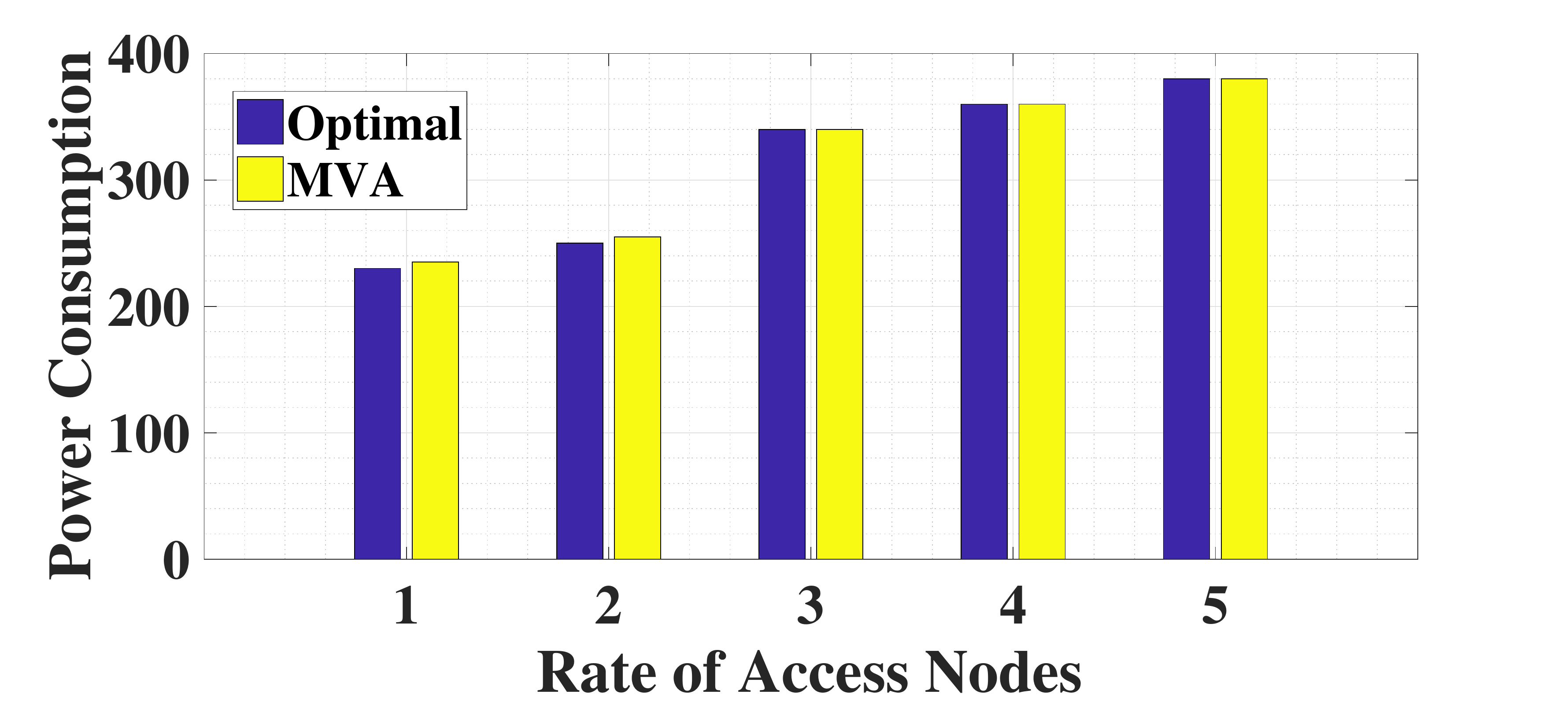}
		\vspace{-.25in}
		\renewcommand{\captionfont}{\small}
		\caption{ Structure 1}
		\label{ structure1}
	\end{subfigure}
	\begin{subfigure}[b]{0.32\textwidth}
		\includegraphics[width=\textwidth]{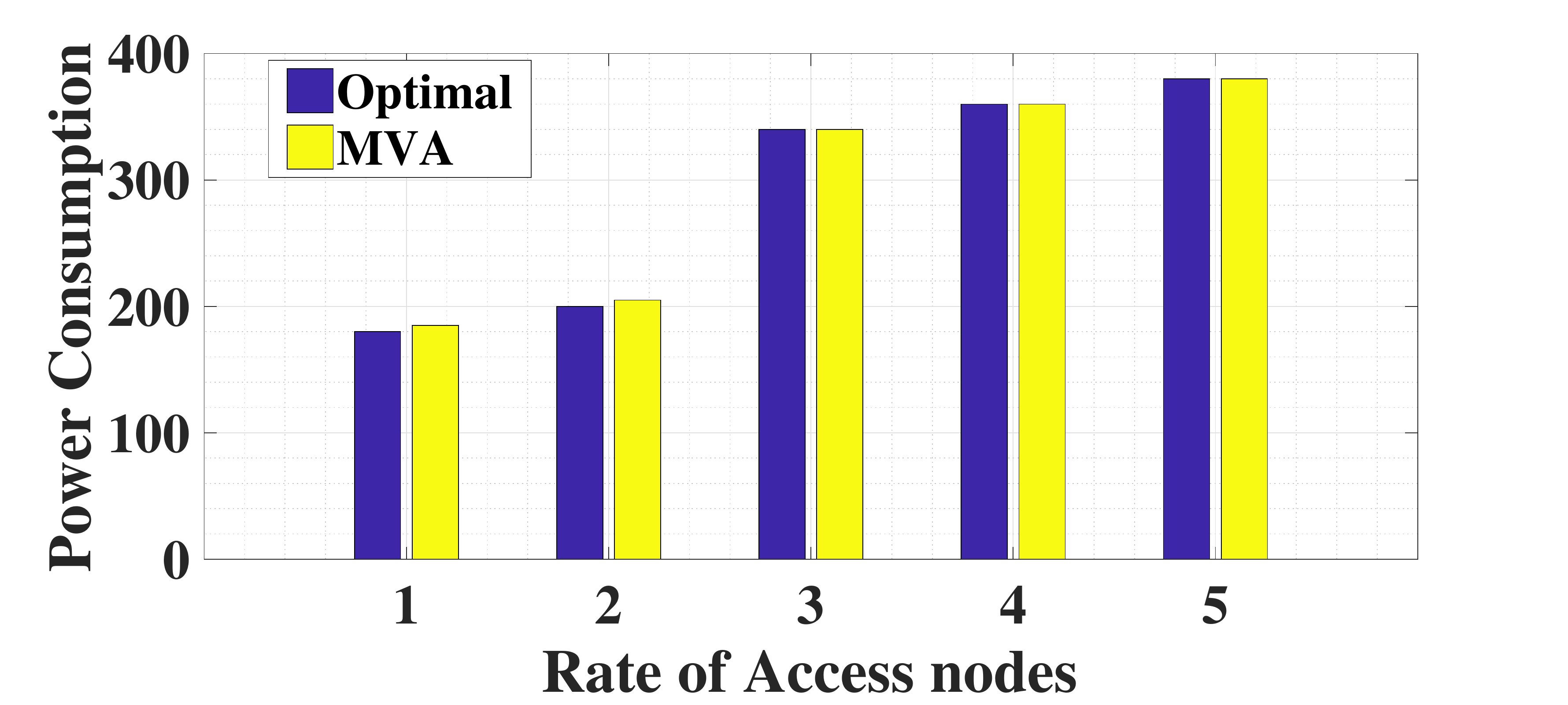}
		\vspace{-.25in}
		\renewcommand{\captionfont}{\small}
		\caption{ Structure 2}
		\label{structure2}
	\end{subfigure}
	\begin{subfigure}[b]{0.32\textwidth}
		\includegraphics[width=\textwidth]{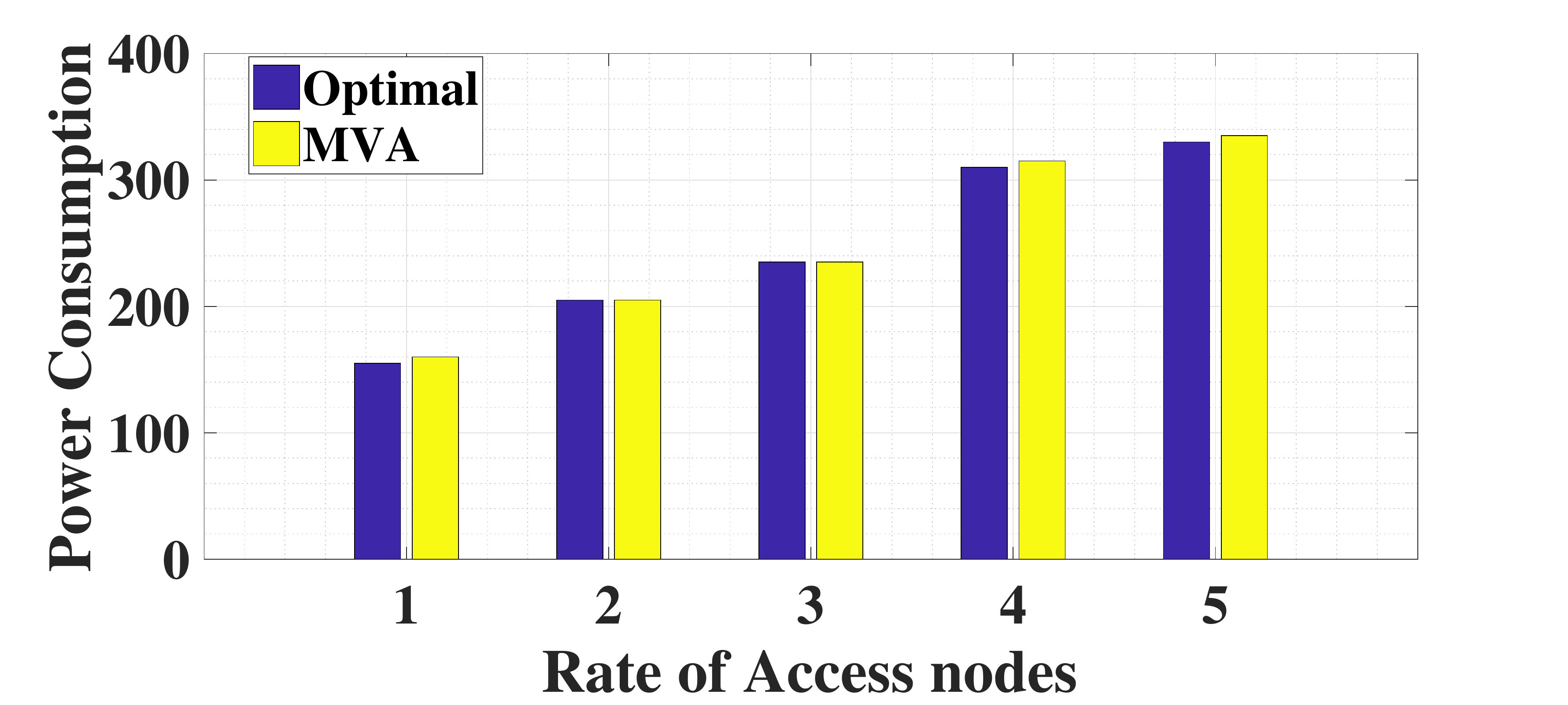}
		\vspace{-.25in}
		\renewcommand{\captionfont}{\small}
		\caption{ Structure 3}
		\label{structure3}
	\end{subfigure}
	\begin{subfigure}[b]{0.32\textwidth}
		\includegraphics[width=\textwidth]{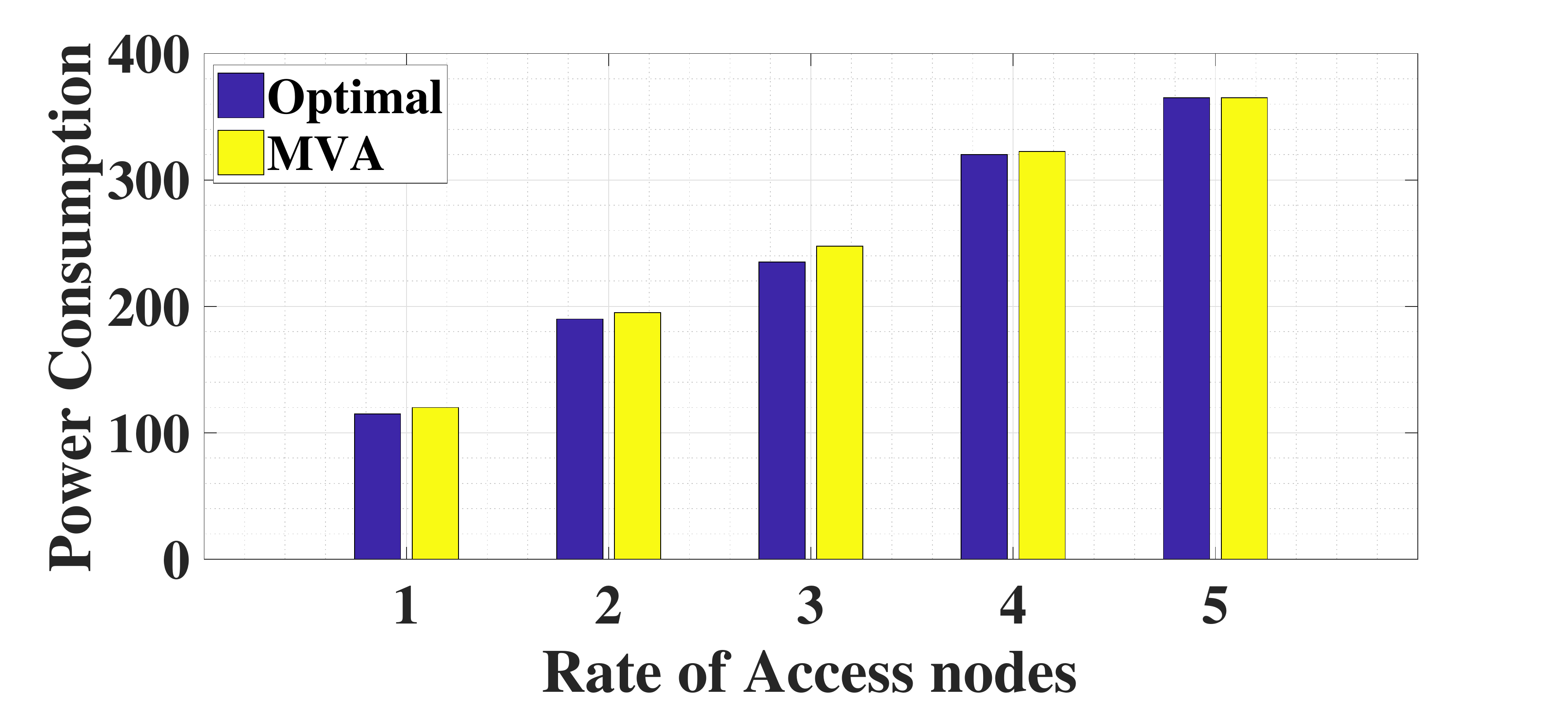}
		\vspace{-.25in}
		\renewcommand{\captionfont}{\small}
		\caption{ Structure 4}
		\label{structure4}
	\end{subfigure}
	\begin{subfigure}[b]{0.32\textwidth}
		\includegraphics[width=\textwidth]{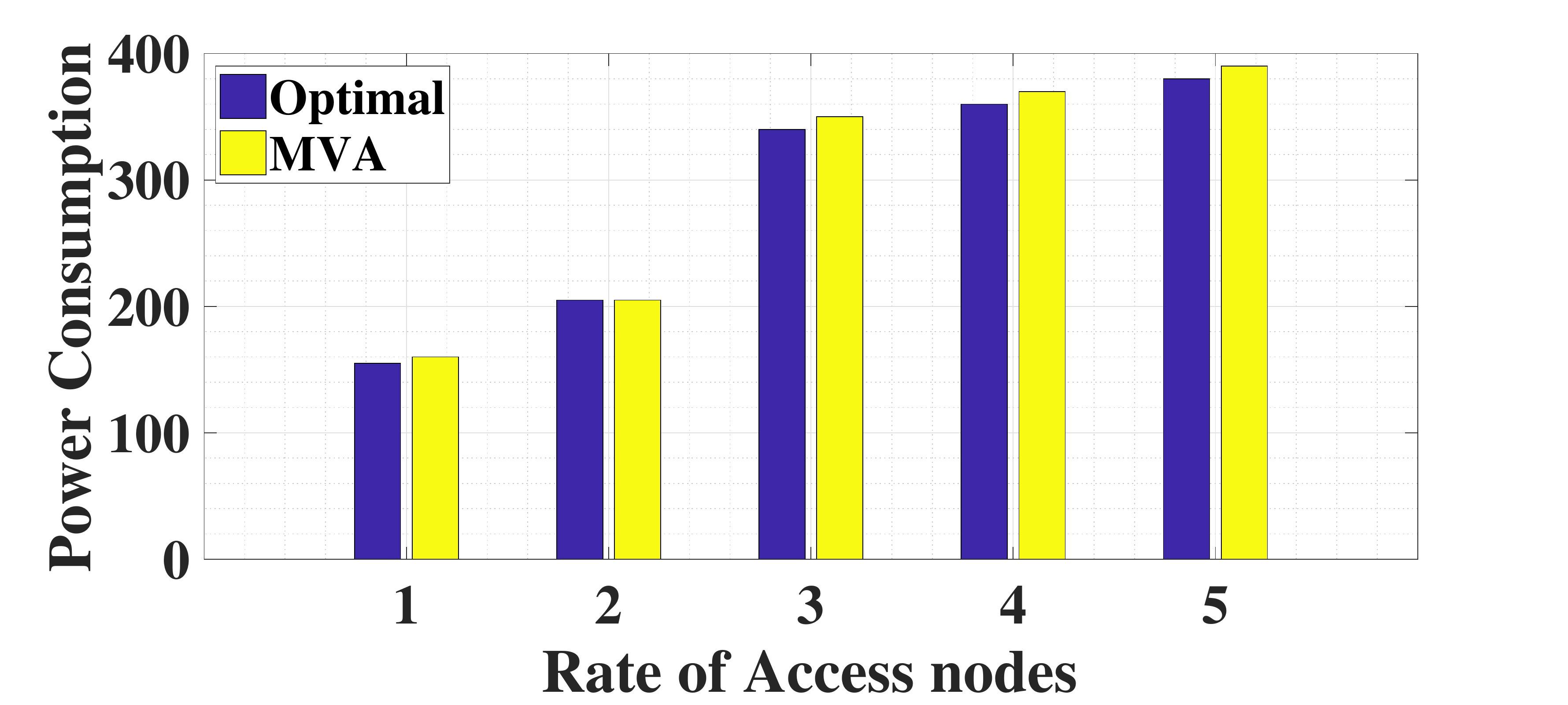}
		\vspace{-.25in}
		\renewcommand{\captionfont}{\small}
		\caption{ Structure 5}
		\label{structure5}
	\end{subfigure}
	\begin{subfigure}[b]{0.32\textwidth}
		\includegraphics[width=\textwidth]{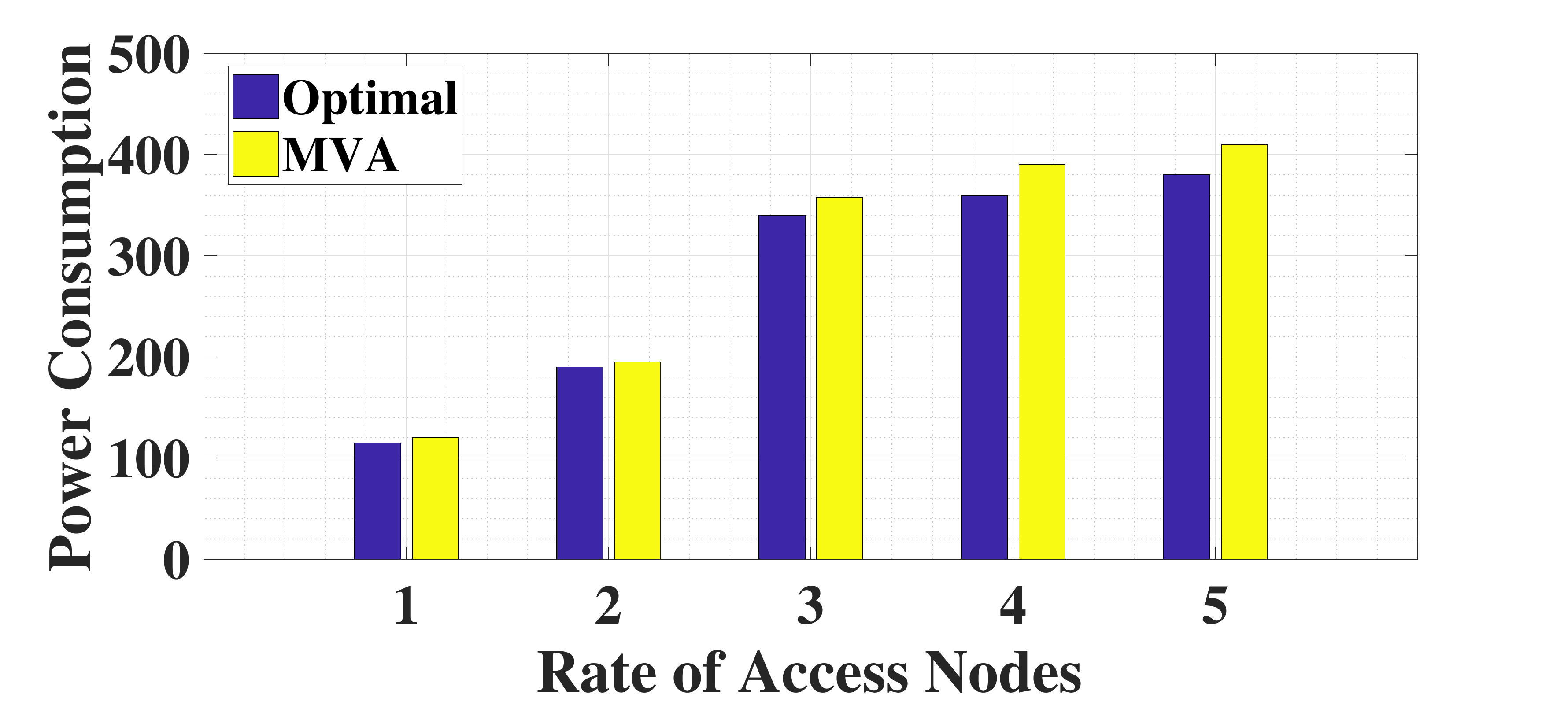}
		\vspace{-.25in}
		\renewcommand{\captionfont}{\small}
		\caption{ Structure 6}
		\label{structure6}
	\end{subfigure}
	\renewcommand{\captionfont}{\small}
	\vspace{-0.2in}
	\caption{Rate of Access Nodes versus power consumption for Optimal Solution and MVA} \label{structures}
	\vspace{-0.2in}
\end{figure*}

To compare  MVA and the optimal solutions, we consider a connected network with 11 nodes and 19 links.
We assume that all switches are SDN. For the nodes that execute the functions, six structures are considered, whose  details  are shown in Table \ref{Sample1}.
The properties of the nodes, links, and functions are shown in Table \ref{Sample2}. For NFV nodes, a resource type is considered, and the capacity of this type of resource is shown in this table. In rows 8 and 9, three numbers are related to NFV node type 1,  NFV node type 2, and  NFV node type 3, respectively.
Two access nodes are considered in the network. We assume that the traffic of access nodes varies from 1 to 5. Figure \ref{structures} shows the solution of the proposed algorithm and the optimal solution (exhaustive search) for different network structures. The figure demonstrates that the proposed algorithm, in most cases, achieves the optimal solution and, in other cases a near-optimal solution.
When there are only NFV nodes or non-NFV (structures 1 to 4), the algorithm obtains the optimal solution. In hybrid NFV structures (structures 5 and 6), the algorithm deviates slightly from the optimal response by increasing the rate of access nodes. However, in the worst case, the proposed algorithm has a response time of 1.1 times the optimal response.
in low- and medium-load cases, the use of NFV nodes always reduces network energy. Conversely, in high-load cases, the consumed energy of the network is not much different from the case where NFV and non-NFV nodes are used.

\begin{table*}
	\vspace{-.3in}
	\caption{Characteristics of Structures}
	\label{Sample1}
	\vspace{-.1in}
	\centering
	\begin{tabular}{lcccc}

		\hline
		%	\rowcolor{gray}\multicolumn{2}{|c|}{\bf Small Size Network}\\
		\rowcolor{gray} Structures & NFV Nodes & Non-NFV Nodes & SDN Nodes & Access Nodes \\

		\hline
		Structure 1                & $0$       & $2$           & $7$       & $2$          \\
		Structure 2                & $2$       & $0$           & $7$       & $2$          \\
		Structure 3                & $4$       & $0$           & $5$       & $2$          \\
		Structure 4                & $8$       & $0$           & $1$       & $2$          \\
		Structure 5                & $2$       & $2$           & $5$       & $2$          \\
		Structure 6                & $4$       & $2$           & $3$       & $2$          \\

		\hline
	\end{tabular}
\end{table*}

\begin{table*}

	\caption{Characteristics of Nodes, Links and Functions}
	\label{Sample2}
	\vspace{-.1in}
	\centering
	\scalebox{0.85}{
		\begin{tabular}{lp{2.1cm}p{2.1cm}p{2.1cm}ccc}

			\hline
			%	\rowcolor{gray}\multicolumn{2}{|c|}{\bf Small Size Network}\\
			\rowcolor{gray} Parameters & \scalebox{0.8}{NFV Node type 1   (Structure 2 and 5)} & \scalebox{0.8}{ NFV Node type 2 (Structure 3 and 6)} & \scalebox{0.8}{NFV Node type 3 (Structure 4)} & Non-NFV Nodes                      & Switches                 & Links \\

			\hline
			Idle Power                 & $50$                                                  & $25$                                                 & $12.5$                                        & $50$                               & $10$                     & $5$   \\
			Peak Power                 & $50$                                                  & $25$                                                 & $12.5$                                        & $50$                               & $0$                      & $0$   \\
			Ingress Capacity           & $10$                                                  & $5$                                                  & $2.5$                                         & $10$                               & -                        & $5$   \\
			Resources                  & $60$                                                  & $30$                                                 & $15$                                          & -                                  & -                        & -     \\
			Functions                  & $1-5$                                                 & $1-5$                                                & $1-5$                                         & $1-3$ or $4-5$                     & -                        & -     \\
			\hline
			\cellcolor{gray} Parameter & \cellcolor{gray} VNF 1                                & \cellcolor{gray}  VNF  2                             & \cellcolor{gray} VNF  3                       & \cellcolor{gray}  VNF  4           & \cellcolor{gray}  VNF  5 &       \\
			\cline{1-6}
			Resource Required          & $20$ or $10$ or $5$                                   & $20$ or $10$ or $5$                                  & $20$ or $10$ or $5$                           & $30$ or $15$ or $7.5$              & $30$ or $15$ or $7.5$    &       \\
			Ingress Capacity           & $10$ or $5$ or $2.5$                                  & $10$ or $5$ or $2.5$                                 & $10$ or $5$ or $2.5$                          & 10 or 5 or 2.5$10$ or $5$ or $2.5$ & $10$ or $5$ or $2.5$     &       \\
			\cline{1-6}
		\end{tabular}
	}
	\vspace{-0.3in}
\end{table*}

Three MCNs are based on LTE coverage map in \cite{map}: 1) \textit{a small-sized network} which can be used for Poland, 2) \textit{a medium-sized network} which can be deployed to model regions like  Iran, and 3) \textit{a large-sized network} which is according to the USA network model. The parameters of these networks are displayed in Table \ref{table4}. The number of nodes in the networks and the data rates of flows are based on \cite{enhancingLakshman}. In this table, the access nodes generate traffic flows.

The characteristics of the network nodes and NFs are shown in Table \ref{table2}.  The types of switches are in accordance with the setup in \cite{boutaba}.
The traffic rate for each flow is a random variable uniformly chosen between 1Mbps and 900Mbps, and we consider a set of NFs as $\mathcal{G}=\{g_1,g_2,g_3,g_4,g_5\}$. Assume there exist two types of non-NFV nodes: 1) SGW $\mathcal{G}_{u1}=\{g_1,g_2,g_3\}$ and, 2) PGW $\mathcal{G}_{u2}=\{g_4,g_5\}$. We suppose that all NFs must be executed for all flows, i.e., $\mathcal{G}=\mathcal{G}^{f}$ for all $f$.
Since there exists no related work to this paper, we compare our results with the case that only one path is stored in each stage, i.e., the traditional VA.
\begin{figure*}
	\vspace{-0.1in}
	\begingroup\makeatletter\def\f@size{11}\check@mathfonts
		\begin{equation}\label{Define_e}
			\eta=\frac{\text{Consumed power of  network using MVA}}{\text{ Consumed  power of reference network while all its network components are active }}
		\end{equation}
		\hrule
	\endgroup
\end{figure*}

\begin{table*}

	\vspace{-.1in}
	\caption{Parameters of three scenarios }
	\label{table4}
	\vspace{-.1in}
	\centering
	\begin{tabular}{lccc}

		\hline
		%	\rowcolor{gray}\multicolumn{2}{|c|}{\bf Small Size Network}\\
		\rowcolor{gray} Parameters               & Small-Sized Network & Medium-Sized Network & Large-Sized Network \\

		\hline
		Number of Access Nodes                   & $16$                & $60$                 & $100$               \\
		Number of Switches                       & $32$                & $90$                 & $150$               \\
		Number of Links                          & $88$                & $282$                & $460$               \\
		Number of Backbone Links                 & $4$                 & $42$                 & $60$                \\
		Number of Type 1 of Non-NFV Nodes (SGW)  & $4$                 & $6$                  & $10$                \\
		Number of Type 2 of Non-NFV Nodes (PGW)  & $2$                 & $3$                  & $5$                 \\
		Number of NFV Nodes                      & $8$                 & $12$                 & $25$                \\
		Number of Internet Exchange Points (IxP) & $1$                 & $3$                  & $5$                 \\
		Capacity of links                        & $[1Mbps,1Gbps]$     & $[1Mbps,1Gbps]$      & $[1Mbps,1Gbps]$     \\
		Capacity of backbone links               & $40Gbps$            & $40Gbps$             & $40Gbps$            \\
		Data rate of flow of access nodes        & $[1Mbps,900Mbps]$   & $[1Mbps,900Mbps]$    & $[1Mbps,900Mbps]$   \\

		\hline
	\end{tabular}
	\vspace{-0.3in}
\end{table*}

In all simulation scenarios, the consumed power in our algorithm is compared to the result of the case where all network components are active. When the number of network components increases or decreases, the consumed power is compared to the reference networks, e.g., the networks in Table \ref{table4}. Therefore, we define the power consumption coefficient as \eqref{Define_e}.

\begin{table*}
	\vspace{-.2in}
	\caption{The characteristics of the network nodes and network functions}
	\label{table2}
	\vspace{-0.1in}
	\centering
	\scalebox{0.8}{
		\begin{tabular}{c l c c c c c }
			\hline
			\multicolumn{1}{ c  }{\multirow{6}{*}{\rotatebox{90}{\scalebox{0.9}{{\bf    Network Nodes}} }}} & \cellcolor{gray} Parameter                                   & \cellcolor{gray} NFV Nodes & \cellcolor{gray} Switches Nodes & \cellcolor{gray} \scalebox{0.8}{Type 1 of Non-NFV Nodes (SGW)} & \cellcolor{gray} \scalebox{0.8}{Type 2 of Non-NFV Nodes (PGW)} & \cellcolor{gray}           \\
			\cline{2-7}
			\multicolumn{1}{ c  }{}                                                                         & \cellcolor{lightgray} Idle Power ($\theta P_u^{\text{max}}$) & $1000W$                    & $500W$ (Chassis Power)          & $8000W$                                                        & $8000W$                                                        &                            \\
			\multicolumn{1}{ c  }{}                                                                         & \cellcolor{lightgray} Peak Power ($ P_u^{\text{max}}$)       & $2000W$                    & $1000W$ (Line-cards Power)      & $20000W$                                                       & $20000W$                                                       &                            \\
			\multicolumn{1}{ c  }{}                                                                         & \cellcolor{lightgray} Ingress Capacity ($r_u$)               & ---                        & ---                             & $10Gbps$                                                       & $20Gbps$                                                       &                            \\
			\multicolumn{1}{ c  }{}                                                                         & \cellcolor{lightgray} Functions                              & ---                        & ---                             & $g_1, g_2,g_3$                                                 & $g_4, g_5$                                                     &                            \\
			\multicolumn{1}{ c  }{}                                                                         & \cellcolor{lightgray} Physical CPU Cores                     & 16                         & ---                             & ---                                                            & ---                                                            &                            \\
			\hline
			\multicolumn{1}{ c }{\multirow{4}{*}{\rotatebox{90}{{\bf  \scalebox{1}{NFs}} }}}                & \cellcolor{gray} Parameter                                   & \cellcolor{gray} VNF $g_1$ & \cellcolor{gray} VNF $g_2$      & \cellcolor{gray}?VNF $g_3$                                     & \cellcolor{gray} VNF $g_4$                                     & \cellcolor{gray} VNF $g_5$ \\
			\cline{2-7}
			\multicolumn{1}{ c  }{}                                                                         & \cellcolor{lightgray}CPU Required ($ c_k$)                   & $2$                        & $6$                             & $4$                                                            & $4$                                                            & $8$                        \\
			\multicolumn{1}{ c  }{}                                                                         & \cellcolor{lightgray}Ingress Capacity ($r_j$)                & $1Gbps$                    & $1Gbps$                         & $1Gbps$                                                        & $1Gbps$                                                        & $1Gbps$                    \\
			\multicolumn{1}{ c  }{}                                                                         & \cellcolor{lightgray}Data Rising Factor ($\gamma_k$)         & $1$                        & $1.1$                           & $1$                                                            & $1$                                                            & $1.05$                     \\
			\hline
		\end{tabular}
	}
	\vspace{-0.1in}
\end{table*}

\begin{figure*}
	\centering
	\begin{subfigure}[b]{0.32\textwidth}
		\includegraphics[width=\textwidth]{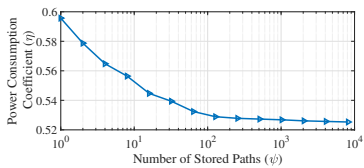}
		\vspace{-.25in}
		\renewcommand{\captionfont}{\small}
		\caption{Small-sized network}
		\label{savesmall}
	\end{subfigure}
	\begin{subfigure}[b]{0.32\textwidth}
		\includegraphics[width=\textwidth]{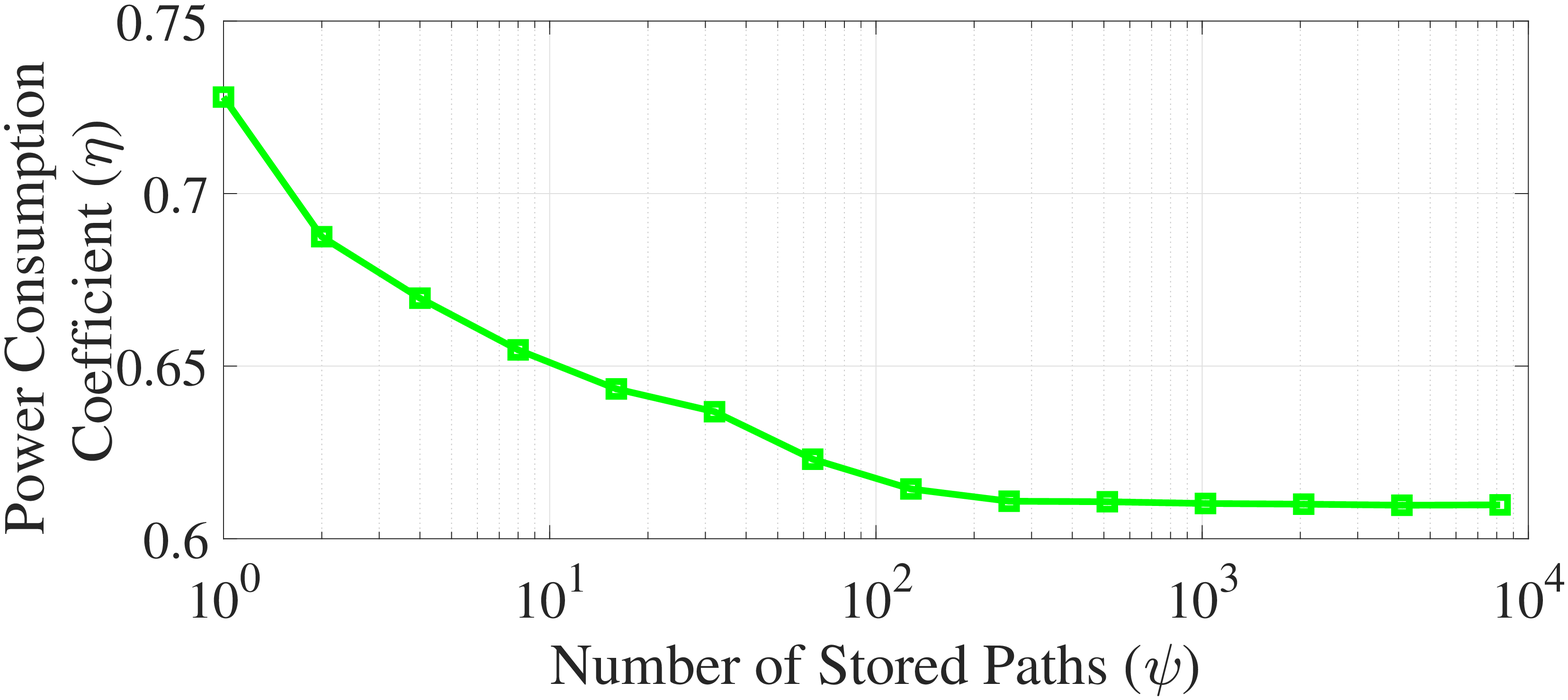}
		\vspace{-.25in}
		\renewcommand{\captionfont}{\small}
		\caption{Medium-sized network}
		\label{savemedium}
	\end{subfigure}
	\begin{subfigure}[b]{0.32\textwidth}
		\includegraphics[width=\textwidth]{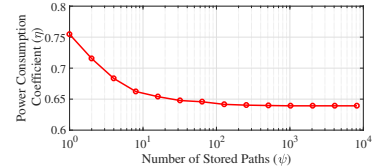}
		\vspace{-.25in}
		\renewcommand{\captionfont}{\small}
		\caption{Large-sized network}
		\label{savelarge}
	\end{subfigure}
	\vspace{-.15in}
	\renewcommand{\captionfont}{\small}
	\caption{Number of stored paths ($\psi$) versus the power consumption coefficient ($\eta$)} \label{save}
	\vspace{-0.25in}
\end{figure*}

One of the important parameters to improve the performance of the proposed MVA is the number of stored paths, i.e.,  $\psi^f_{\bar{k}}$. To show the effect of this parameter, we evaluate the amount of power consumed versus $\psi^f_{\bar{k}}$. We also consider $\psi=\psi^f_{\bar{k}},~\forall \bar{k},~ \forall f$ and that all switches are SDN. In Fig. \ref{save}, the number of stored paths ($\psi$) versus the power consumption coefficient  ($\eta$) is demonstrated for three network scenarios. As expected, it reveals from Fig. \ref{save} that storing more paths for each node leads a better energy efficiency in the networks. In small-sized, medium-sized, and large-sized networks, if 64, 128, and 128 paths are stored for each node, the consumed power decrements about 8\%, 10\%, and 12\%, respectively. From Fig. \ref{save}  when $\psi= 1024$, the performance of our algorithm approaches the case that all paths are stored in the three network scenarios.

\begin{figure*}
	\centering
	\vspace{-0.25in}
	\begin{subfigure}[b]{0.32\textwidth}
		\includegraphics[width=\textwidth]{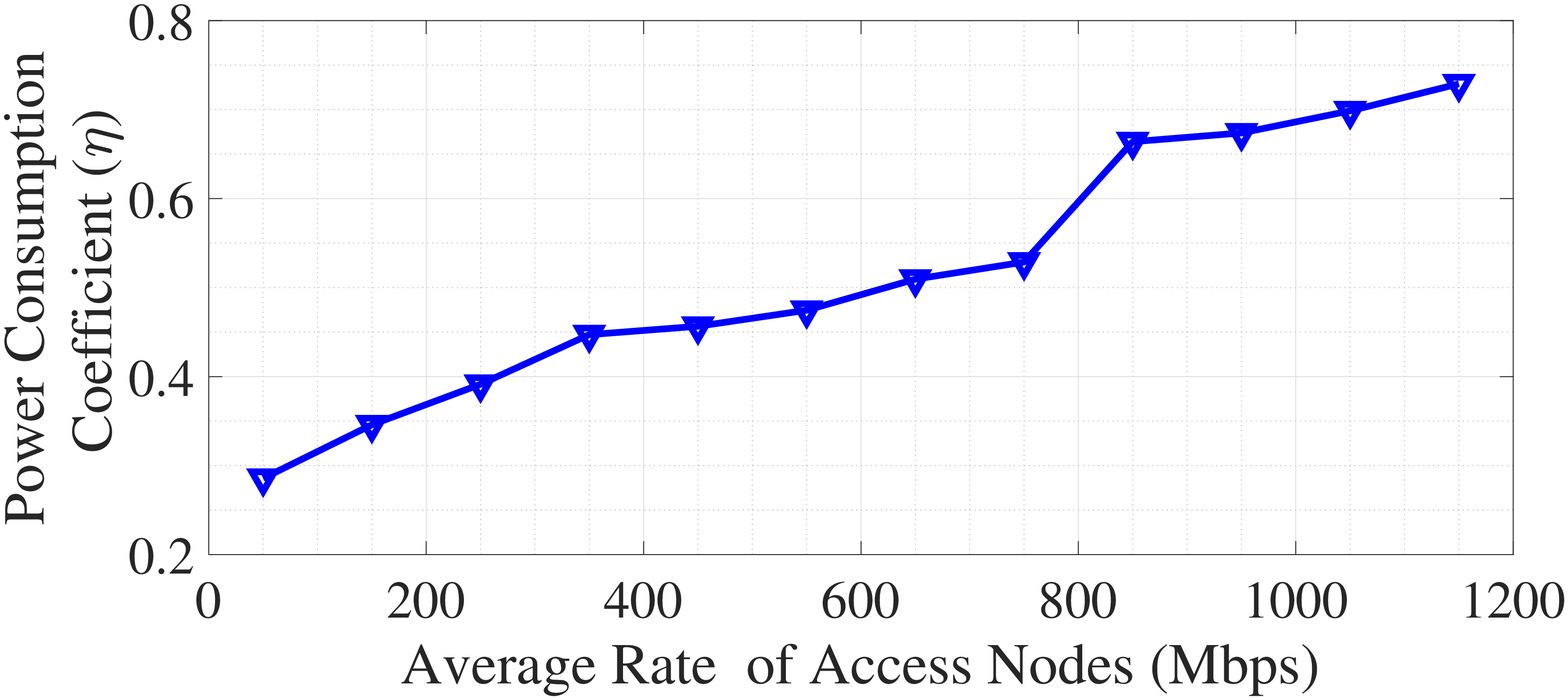}
		\vspace{-.25in}
		\renewcommand{\captionfont}{\small}
		\caption{Small-sized network}
		\label{loadsmall}
	\end{subfigure}
	\begin{subfigure}[b]{0.32\textwidth}
		\includegraphics[width=\textwidth]{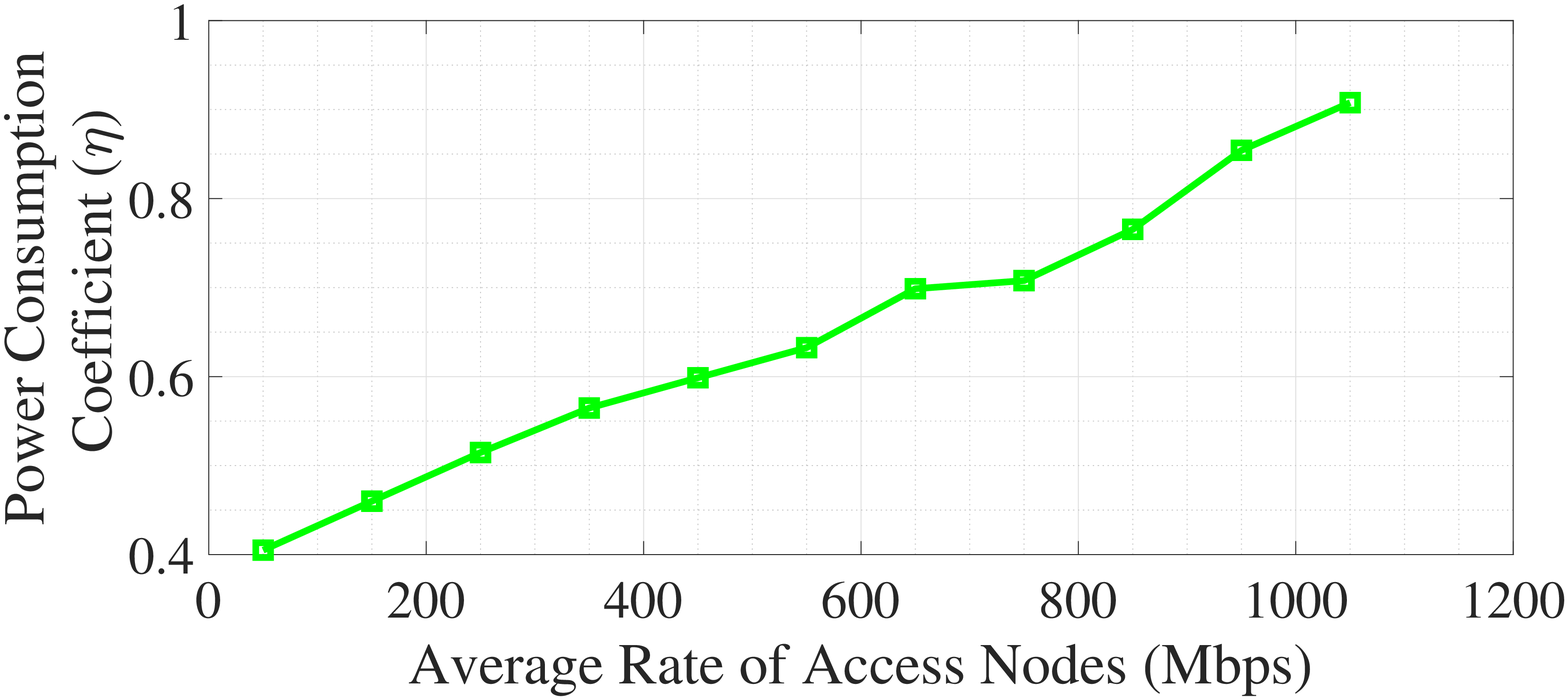}
		\vspace{-.25in}
		\renewcommand{\captionfont}{\small}
		\caption{Medium-sized network}
		\label{loadmedium}
	\end{subfigure}
	\begin{subfigure}[b]{0.32\textwidth}
		\includegraphics[width=\textwidth]{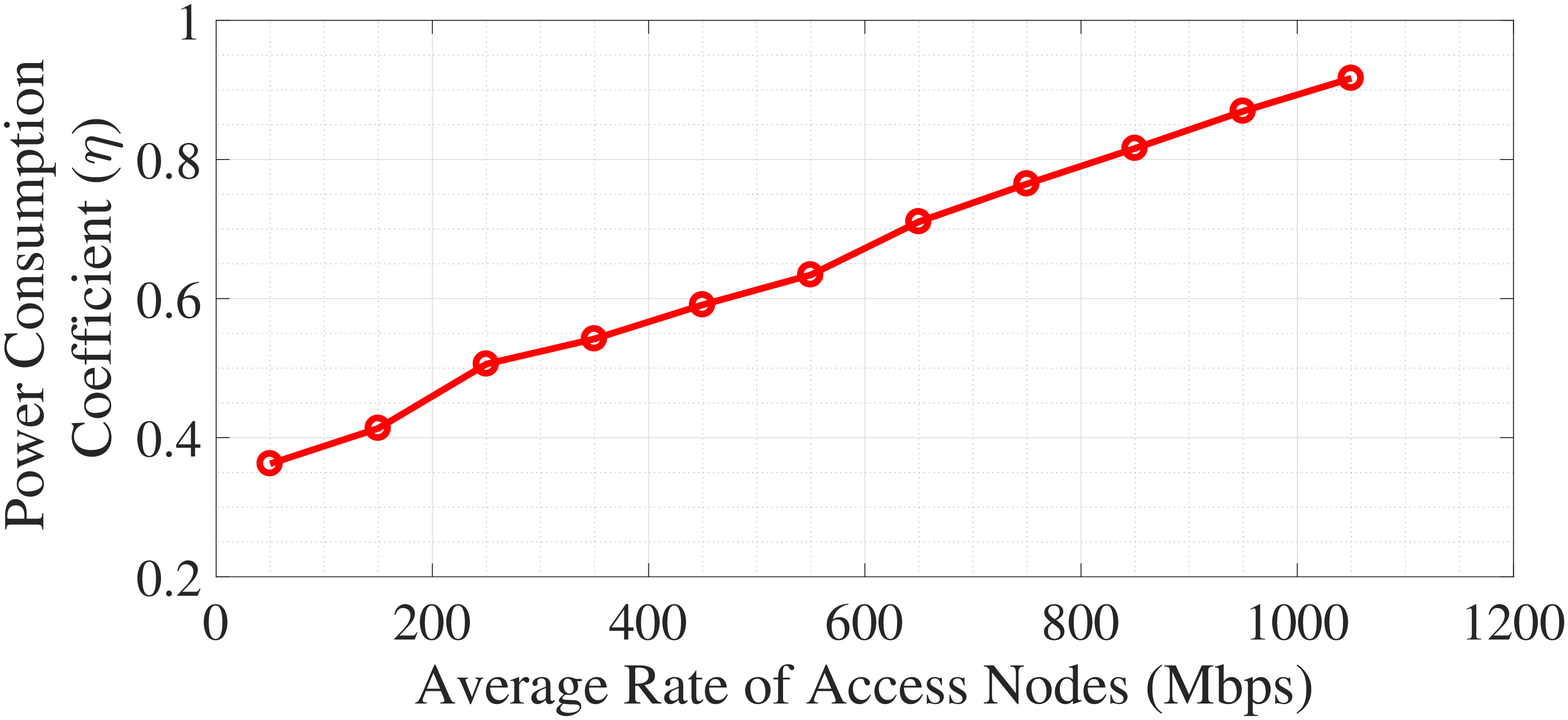}
		\vspace{-.25in}
		\renewcommand{\captionfont}{\small}
		\caption{Large-sized network}
		\label{loadlarge}
	\end{subfigure}
	\vspace{-.2in}
	\renewcommand{\captionfont}{\small}
	\caption{Average rate of flows versus the power consumption coefficient ($\eta$) }\label{load}
	\vspace{-.3in}
\end{figure*}

Fig. \ref{load} plots the percentage of the consumed power in the three network scenarios versus the average data rate of the access nodes, i.e., $r^f$. For this simulation, we assume all switches are SDN. Fig. \ref{load} (a) shows that for the small-sized network, by increasing the average data rate of the access nodes from 50Mbps to 1150Mbps, the consumed power of the network rises from 30\% to 75\%.   Figures \ref{load} (b) and \ref{load} (c) show that in the medium-sized and large-sized networks by increasing the average data rate of access nodes, the consumed power of networks rises approximately from 40\% to 90\%.  These results verify that increasing $r^f$  leads to an increment of the consumed power as expected.

\begin{figure*}
	\centering
	\vspace{-0.3in}
	\begin{subfigure}[b]{0.32\textwidth}
		\includegraphics[width=\textwidth]{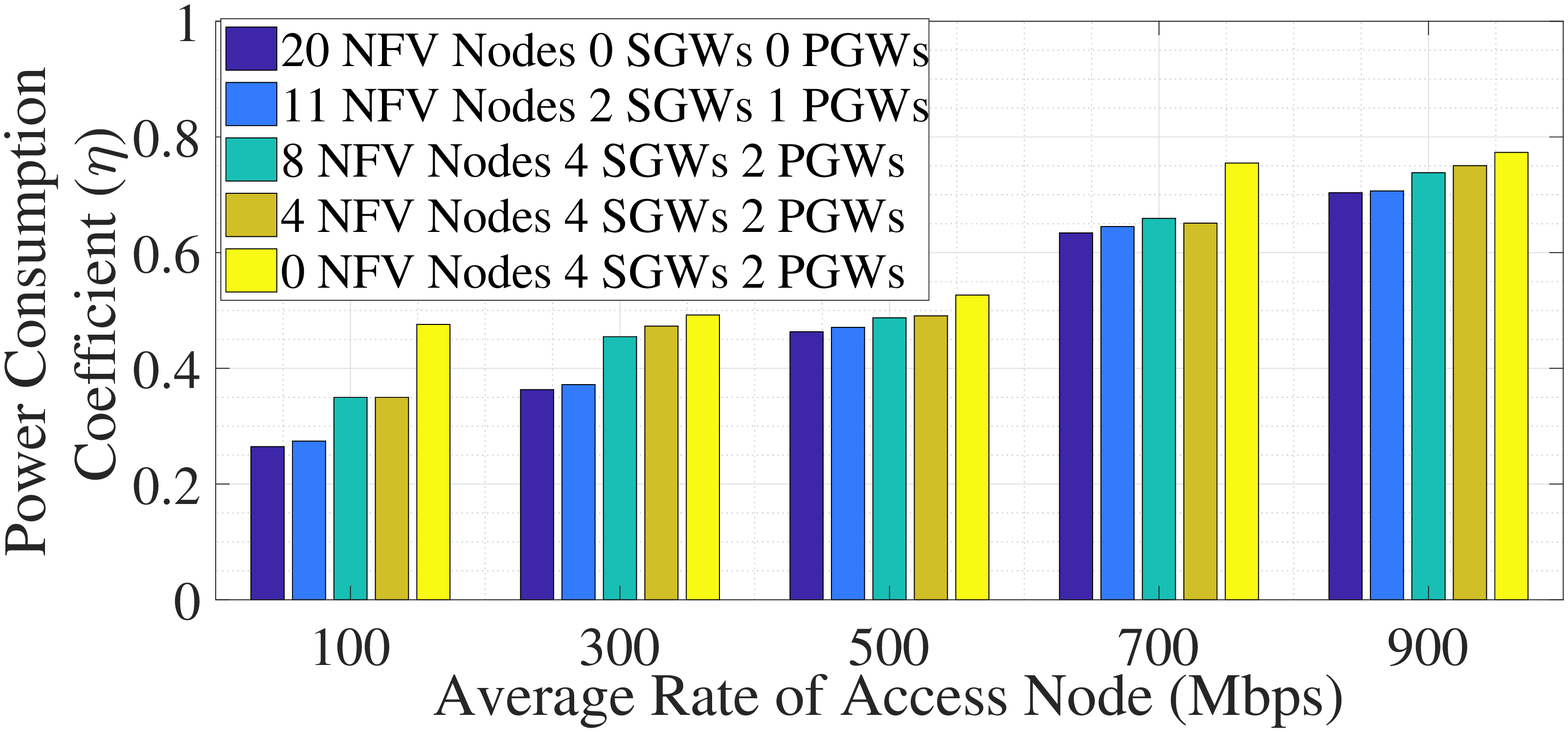}
		\vspace{-.25in}
		\renewcommand{\captionfont}{\small}
		\caption{Small-sized network}
		\label{fullsmall}
	\end{subfigure}
	\begin{subfigure}[b]{0.32\textwidth}
		\includegraphics[width=\textwidth]{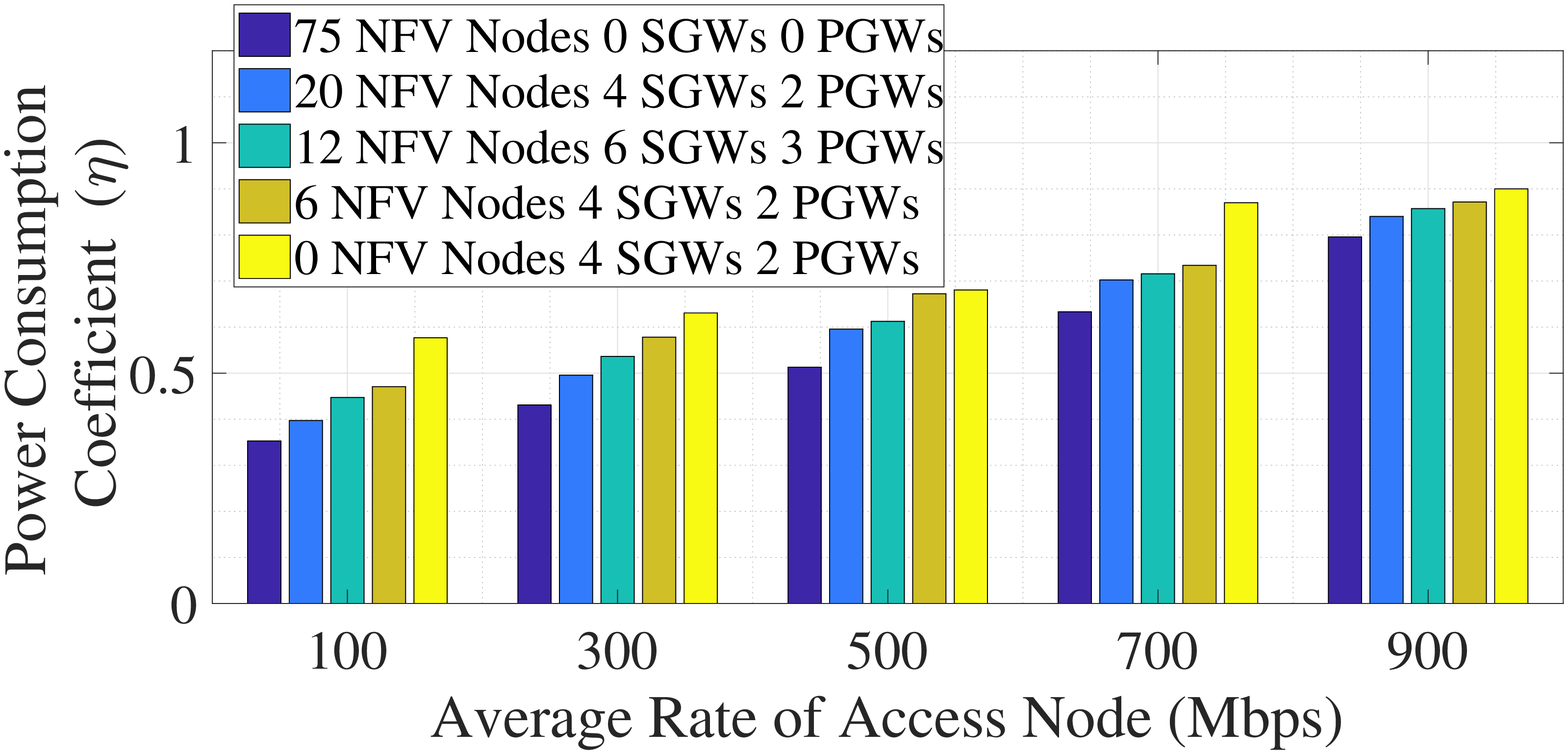}
		\vspace{-.25in}
		\renewcommand{\captionfont}{\small}
		\caption{Medium-sized network}
		\label{fullmedium}
	\end{subfigure}
	\begin{subfigure}[b]{0.32\textwidth}
		\includegraphics[width=\textwidth]{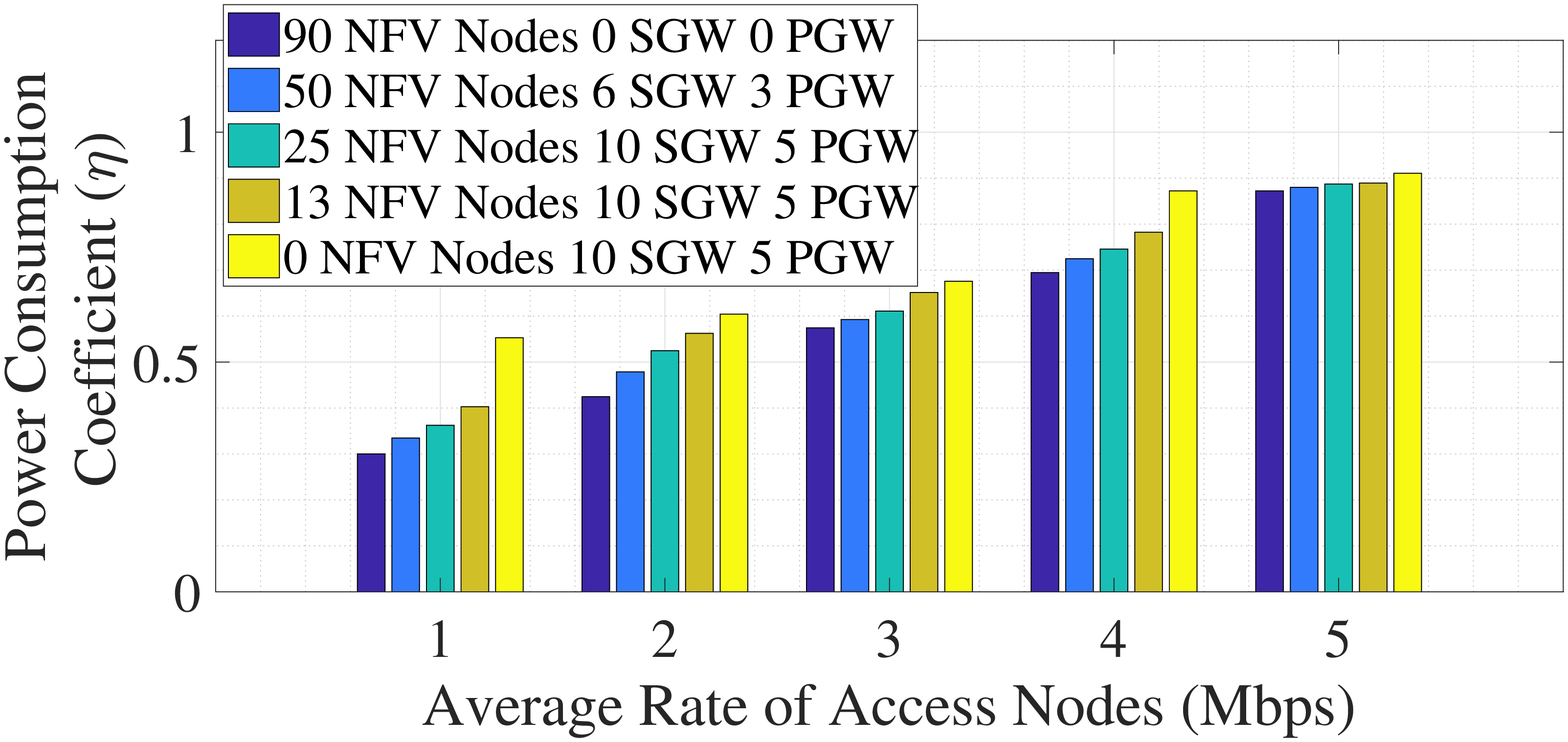}
		\vspace{-.25in}
		\renewcommand{\captionfont}{\small}
		\caption{Large-sized network}

		\label{fulllarge}
	\end{subfigure}
	\vspace{-.2in}
	\renewcommand{\captionfont}{\small}
	\caption{Number of NFV nodes and non-NFV nodes versus the power consumption coefficient ($\eta$) for 100\% SDN switches }\label{full}
	\vspace{-0.1in}
\end{figure*}

\begin{figure*}
	\centering
	\vspace{-0.2in}
	\begin{subfigure}[b]{0.32\textwidth}
		\includegraphics[width=\textwidth]{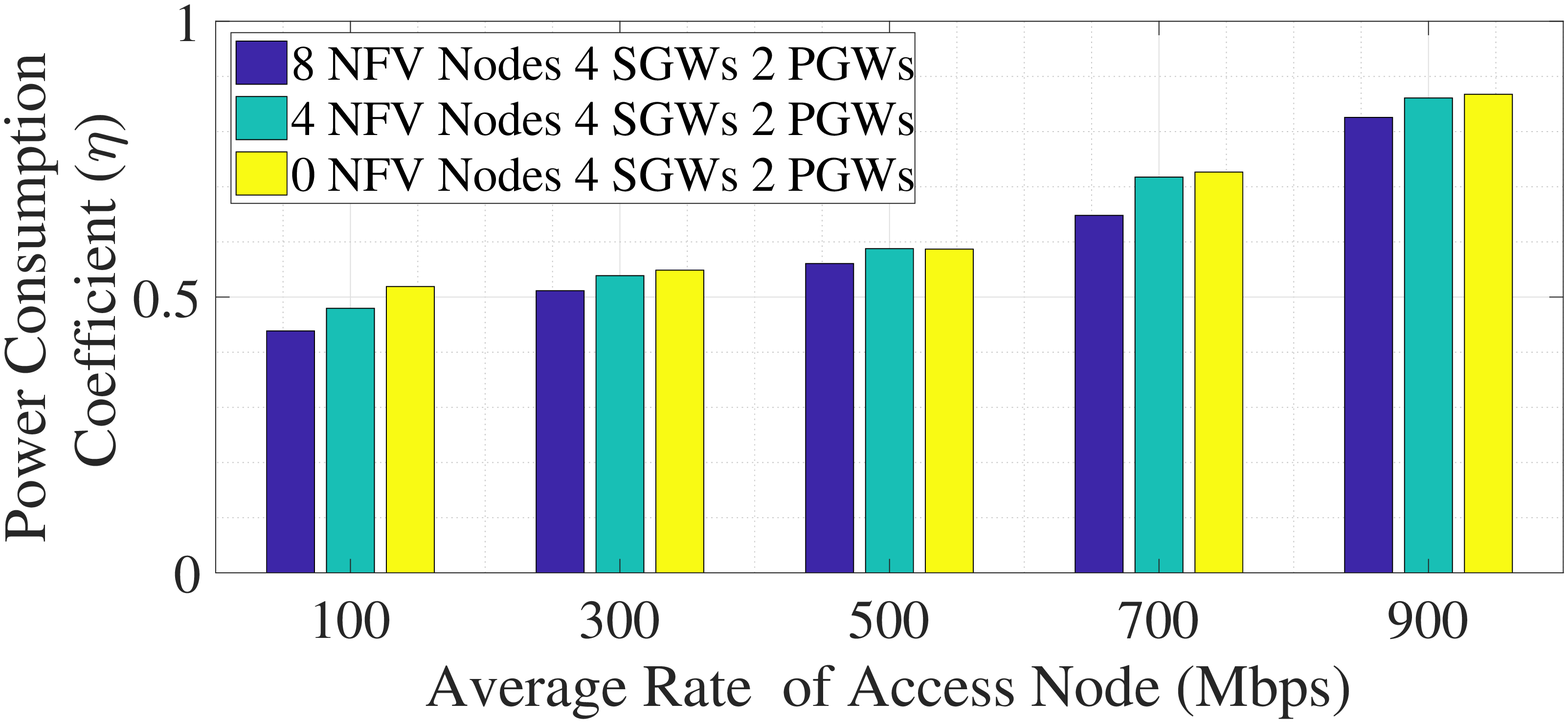}
		\vspace{-.25in}
		\renewcommand{\captionfont}{\small}
		\caption{Small-sized network}
		\label{halfsmall}
	\end{subfigure}
	\begin{subfigure}[b]{0.32\textwidth}
		\includegraphics[width=\textwidth]{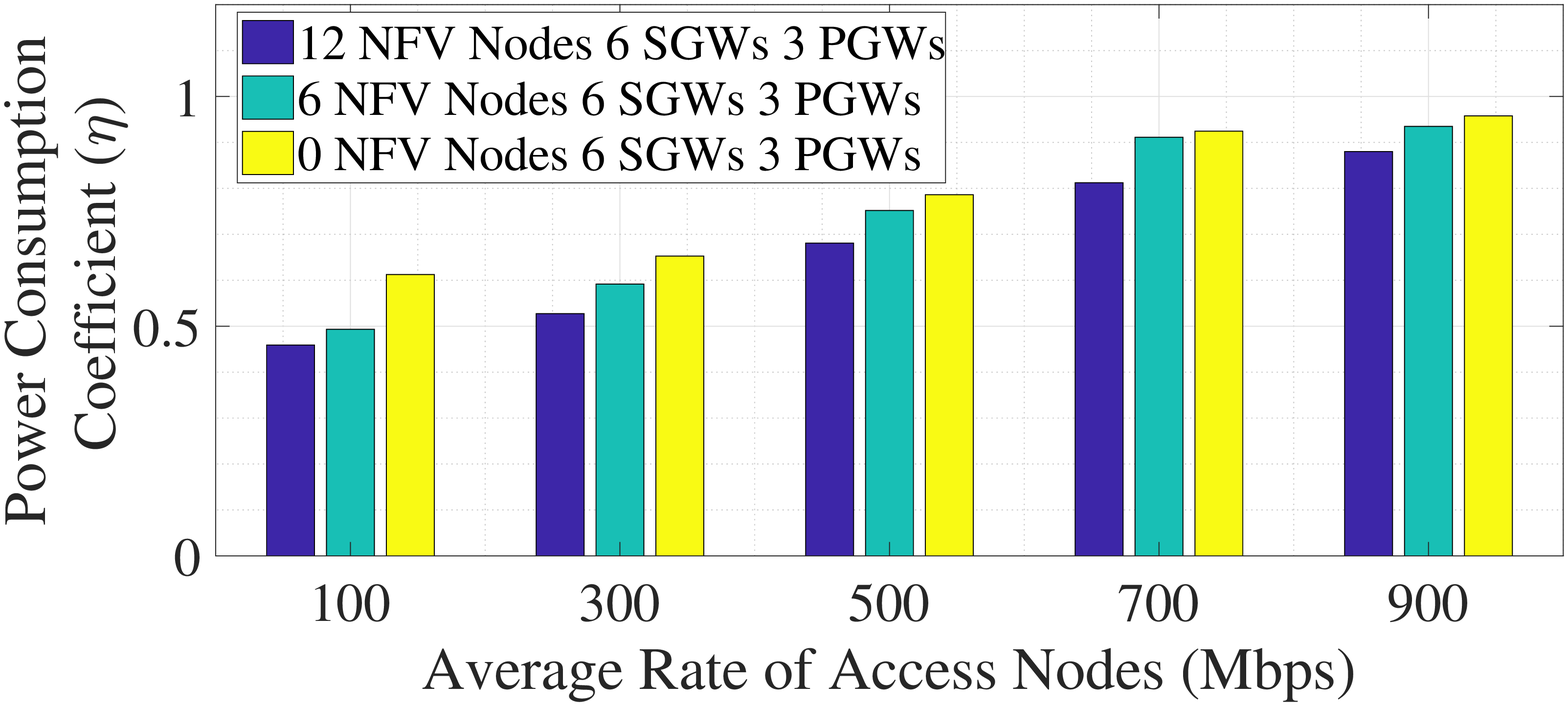}
		\vspace{-.25in}
		\renewcommand{\captionfont}{\small}
		\caption{Medium-sized network}
		\label{halfmedium}
	\end{subfigure}
	\begin{subfigure}[b]{0.32\textwidth}
		\includegraphics[width=\textwidth]{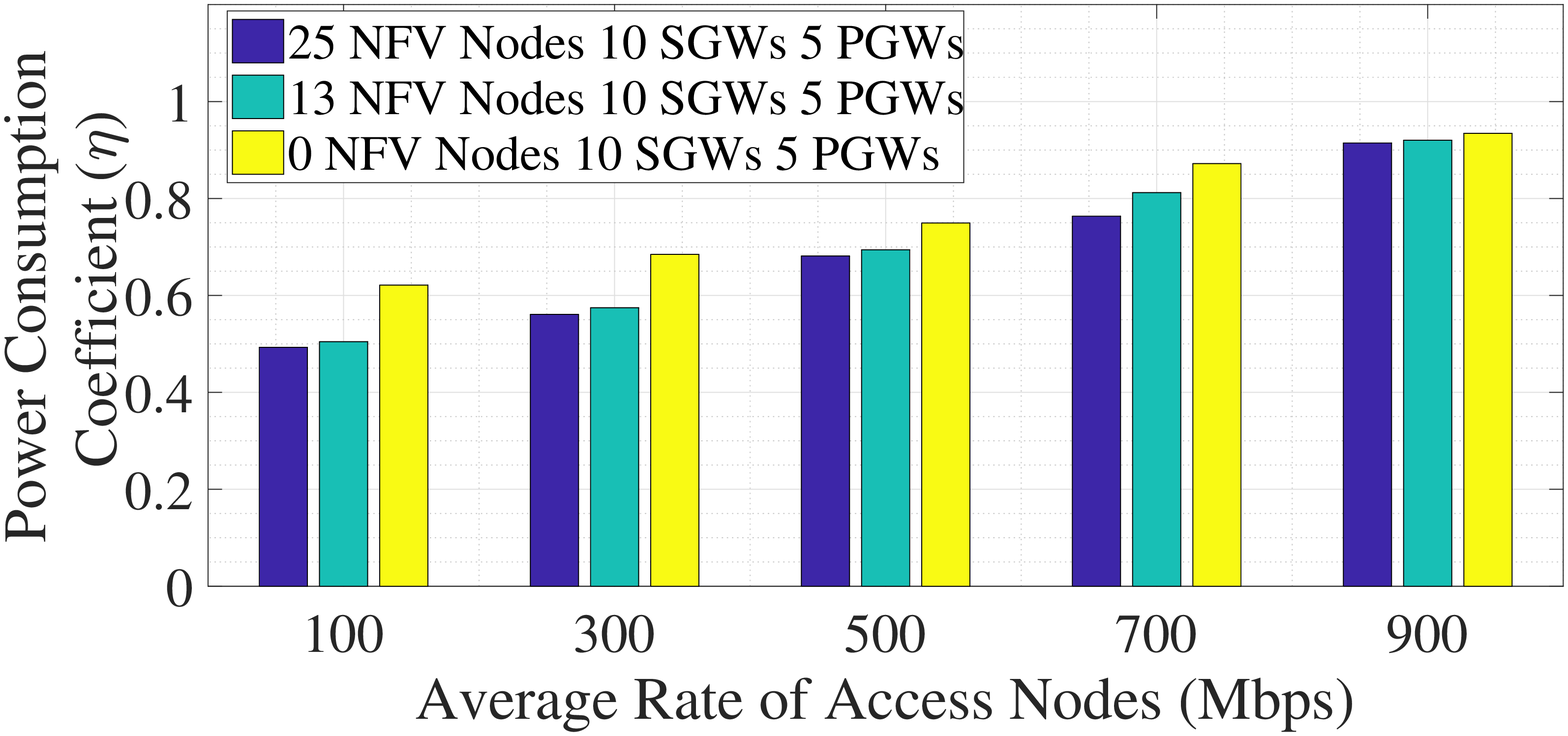}
		\vspace{-.25in}
		\renewcommand{\captionfont}{\small}
		\caption{Large-sized network}

		\label{halflarge}
	\end{subfigure}
	\vspace{-.2in}
	\renewcommand{\captionfont}{\small}
	\caption{Number of NFV nodes and non-NFV nodes versus the power consumption coefficient ($\eta$) for 50\% SDN switches }\label{half}
	\vspace{-0.3in}
\end{figure*}

\begin{figure*}
	\centering
	\vspace{-0.2in}
	\begin{subfigure}[b]{0.32\textwidth}
		\includegraphics[width=\textwidth]{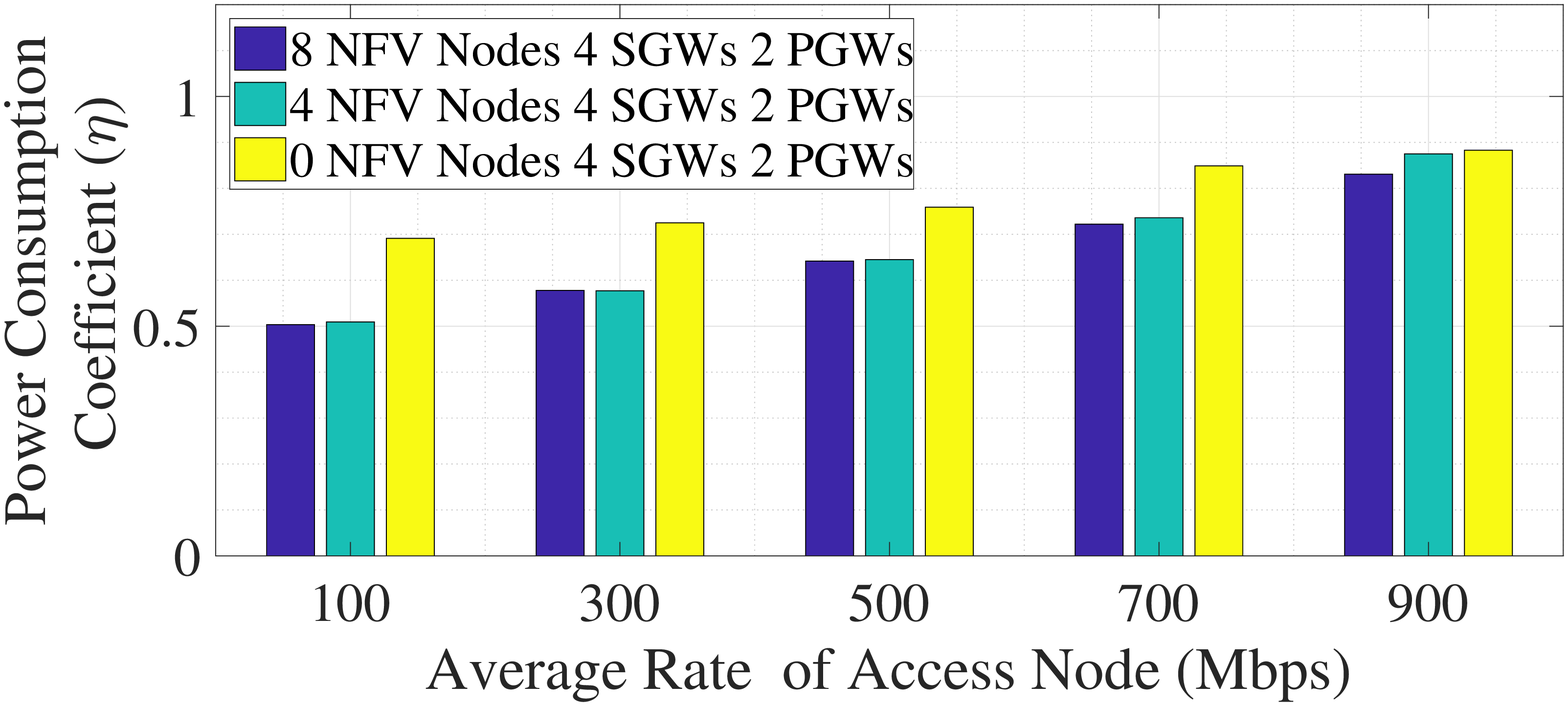}
		\vspace{-.25in}
		\caption{Small-sized network}
		\label{zerosmall}
	\end{subfigure}
	\begin{subfigure}[b]{0.32\textwidth}

		\includegraphics[width=\textwidth]{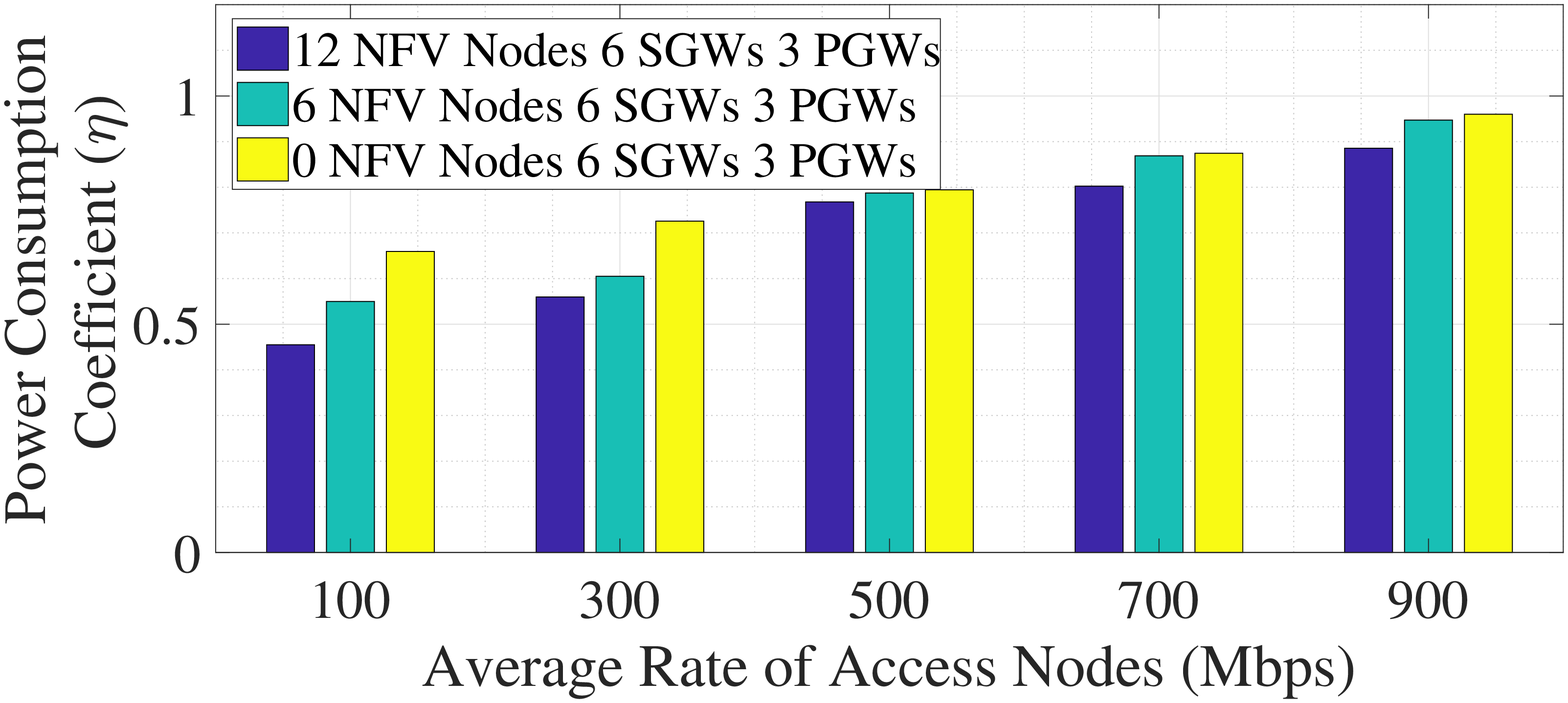}
		\vspace{-.25in}
		\caption{Medium-sized network}
		\label{zeromedium}
	\end{subfigure}
	\begin{subfigure}[b]{0.32\textwidth}
		\includegraphics[width=\textwidth]{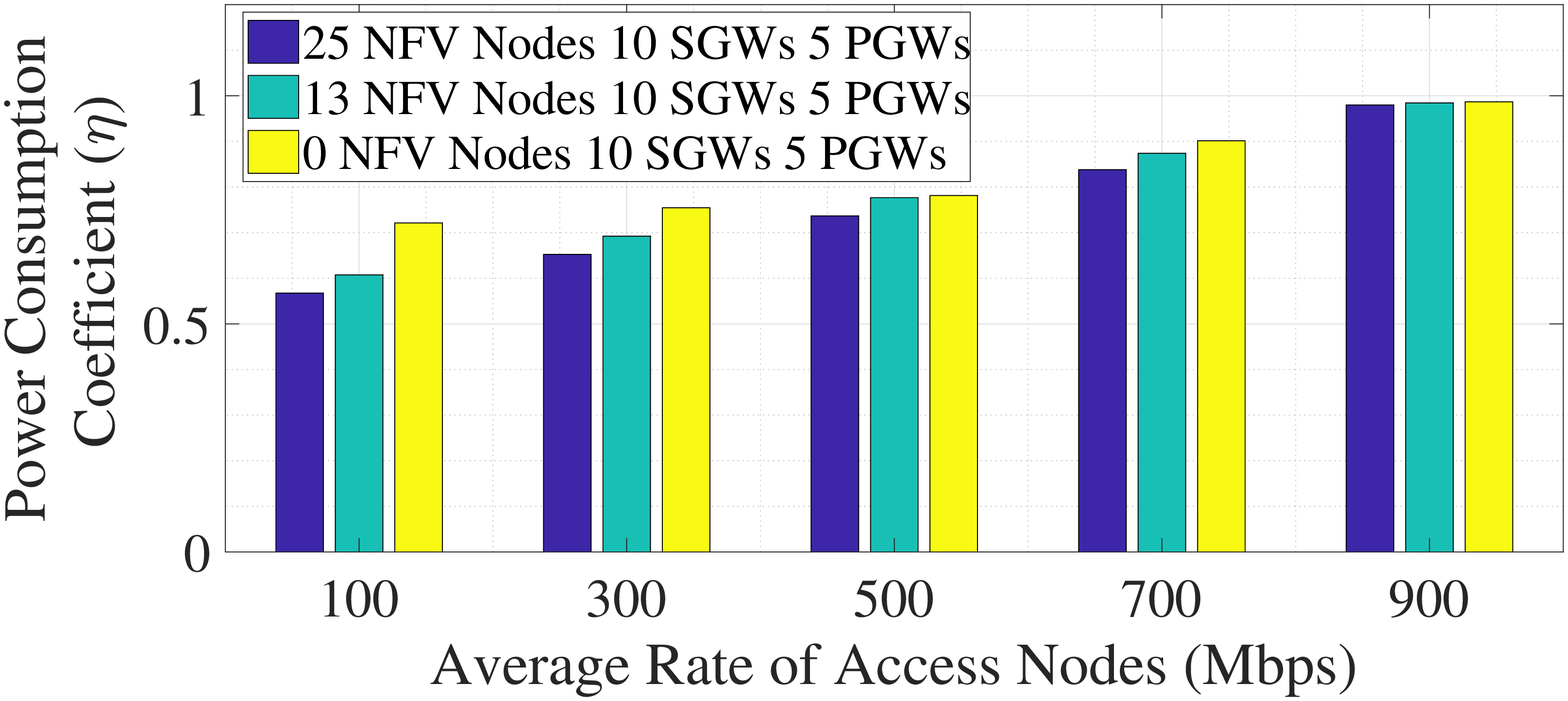}
		\vspace{-.25in}
		\caption{Large-sized network}
		\label{zerolarge}
	\end{subfigure}
	\vspace{-.2in}
	\renewcommand{\captionfont}{\small}
	\caption{Number of NFV nodes and non-NFV nodes versus the power consumption coefficient ($\eta$) for 0\% SDN switches }\label{zero}
	\vspace{-0.3in}
\end{figure*}

Now we aim to evaluate the effect of transition from the non-SDN and non-NFV network to the full SDN and full NFV networks in terms of energy efficiency. To answer this critical issue, in Figs. \ref{full}-\ref{zero}, for different numbers of NFV nodes and non-NFV nodes and for different percentages of SDN switches,  the power consumption coefficient of the three network scenarios are plotted versus the data rate of access nodes. In Fig. \ref{full}, all switches belong to SDN nodes. In Fig. \ref{half}, 50\% of switches belong to SDN nodes, and in Fig. \ref{zero}, all switches are non-SDN. For these figures, the average data rate of the access nodes are 100Mbps, 300Mbps, 500Mbps, 700Mbps, and 900Mbps.

Figure \ref{full} (a) show that by increasing the number of NFV nodes in the small-sized network where all switches of networks are SDN, the consumed power of the network decreases. When there are 20 NFV nodes in the network, the network consumes between 8\% and 20\% less power  than the case that there does not exist any NFV node in the network.
Figure \ref{full} (b) shows that in the medium-sized networks and in the case that there are only NFV nodes in the network, the network consumes about 20\% less power  than in a network without NFV nodes.
In the large-sized network scenario, Fig. \ref{full} (c) shows that by increasing the number of NFV nodes from zero to 90\%, $\eta$ reduces by up to 50\%.

Based on our observation in these simulations, in the case that there are no NFV nodes for the three network scenarios, all nodes are in  on state for a certain amount of traffic  in the network. For example, in the small network scenario in Fig. \ref{load} (a), when the data rate is greater than 500Mbps, all nodes are in the on state. As a result, the consumed power exponentially increases. As concluded  from Fig. \ref{full}, in the case that there is no NFV node, when the average data rate increases from 500Mbps to 700Mbps, the consumed powers of the three network scenarios rise significantly. However, for the hybrid NFV networks, the consumed power will gradually increase by increasing the data rates of access nodes in the network. For the same data rate, with increasing the number of NFV nodes in all three network scenarios, the amount of consumed power decreases.

The same behavior can be observed in Figs. \ref{half} and \ref{zero} for the case that 50\% and 0\% of switches are SDN, respectively, i.e., by increasing the number of NFV nodes in the networks, energy consumption has been reduced. In addition, when there is no NFV node in the networks, by increasing data rates of the access nodes from 500Mbps to 700Mbps, the power consumption coefficient ($\eta$) significantly increases. According to Figs. \ref{full}, \ref{half}, and \ref{zero}, the more number of SDN nodes in the network, the more energy will be saved by increasing the number of the NFV nodes. Similarly, if traffic data rates of the access nodes are low, by increasing the number of NFV nodes, we can save more energy in the network.

According to Figs. \ref{full}, \ref{half}, and \ref{zero}, in most cases, the power consumption coefficient  $\eta$, of the networks with hybrid NFV nodes approach the power of networks with full NFV nodes. For instance, in Fig. \ref{full} (a), $\eta$ for a network with 20 NFV nodes is close to $\eta$ for a network with 11 NFV nodes. According to the migration cost of swapping non-NFV/non-SDN nodes to NFV/SDN node, with the hybrid NFV scenarios, mobile operators can approximately achieve the energy efficiency close to the full NFV scenarios with less cost. Also, in hybrid NFV, the computation of our proposed algorithm is reduced considerably compared to the full NFV case.

\begin{table*}

	\caption{$\bar{\eta}$ for different network sizes and different fractions of NFV nodes and SDN switches}
	\label{table5}
		\vspace{-.1in}

	\scalebox{0.8}{
		\vspace{-.1in}
		\centering
		%\scalebox{0.8}{
		\begin{tabular}{|c|ccccc|ccccc|ccccc|}

			\hline

			%\diaghead{\theadfont Diag ColumnmnHead II}%
			%{Percentage \\ of  SDN switches}{Percentage  of NFV \\ nodes}{SDN Switches (\%)}{NFV Nodes (\%)}

			\multirow{2}{*}{\backslashbox{SDN Switches (\%)}{NFV Nodes (\%)}} & \multicolumn{5}{c|}{Small-Sized Network} & \multicolumn{5}{c|}{Medium-Sized Network} & \multicolumn{5}{c|}{Large-Sized Network}                                                                                       \\
			%\cline{2-16}
			                                                                  & 0\%                                      & 25\%                                      & 50\%                                     & 75\% & 100\% & 0\%  & 25\% & 50\% & 75\% & 100\% & 0\% & 25\% & 50\% & 75\% & 100\% \\
			\hline
			\hline
			\rowcolor{lightgray} 0\%                                          & 1                                        & 1.5                                       & 1.67                                     & --   & --    & 1    & 1.28 & 1.49 & --   & --    & 1   & 1.52 & 1.68 & --   & --    \\
			50\%                                                              & 1.16                                     & 1.67                                      & 1.83                                     & --   & --    & 1.08 & 1.42 & 1.53 & --   & --    & 1.4 & 1.68 & 1.8  & --   & --    \\
			\rowcolor{lightgray} 100\%                                        & 1.67                                     & 2                                         & 2.16                                     & 2.3  & 2.3   & 1.2  & 1.51 & 1.57 & 1.71 & 1.87  & 1.8 & 2.32 & 2.44 & 2.25 & 2.6   \\

			\hline
		\end{tabular}
	}

	\vspace{-0.3in}

\end{table*}

To highlight the effect of the fraction of SDN switches and NFV nodes on energy efficiency among the three scenarios in more clear manner, we define
\begin{equation*}
	\bar{\eta}=\frac{1-\eta^{\text{min}}}{1-\eta_\text{{non-NFV/non-SDN}}^{\text{min}}},
\end{equation*}
where $\eta_\text{{non-NFV/non-SDN}}^{\text{min}}$ denotes the minimum power consumption coefficient of the network in the case of non-SDN and non-NFV. $\eta^{\text{min}}$ is the minimum value of $\eta$. Consequently,
$\bar{\eta}$ demonstrates the percentage of saved power for both cases via applying MVA.
Table \ref{table5} represents $\bar{\eta}$ for different fractions of SDN switches and NFV nodes in the networks in the three scenarios.
According to Table \ref{table5}, our proposed algorithm has a considerable gain in term of energy efficiency in the large-sized and small-sized networks. However, the energy efficiency gain in the medium-sized network is not as much as the two other scenarios.

The MVA algorithm is proposed for the transition to full SDN and full NFV network. According to simulations, partial SDN and hybrid NFV network have more energy efficiency than the network, which does not deploy SDN and NFV.
Since deploying the network with SDN and converting non-NFV nodes to NFV nodes are expensive, by equipping parts of the networks with SDN and converting part of non-NFV nodes to NFV nodes, the proposed algorithm can save considerable power in the network. However, in terms of migration and computation costs, partial SDN and hybrid NFV networks can outperform full SDN and NFV networks. Therefore, appropriately choosing the percentage of SDN and NFV nodes in the network can be considered as an essential design parameter from the infrastructure providers' point of view.

\section{Conclusion}\label{conclusion}
In this paper, the joint virtual network function placement and routing optimization problem with the goal of energy efficiency was addressed in a general framework of a partial SDN and hybrid NFV, where links and nodes can be turned off. We demonstrate how the related problem can be formulated. To deal with the NP-hardness of this problem and solve it efficiently, {we deployed a multi-stage graph and a modified Djikstra's routing algorithm (MDRA) by proposing a modified Viterbi algorithm (MVA)}. Simulation results reveal that for small, medium and large network sizes, the proposed algorithm saves up to 70\%, 60\%, and 60\% consumed power compared to the case where all nodes and links are in on-state. The evaluations also assessed the trade-off of changing the percentage of SDN switches and NFV nodes in the network in terms of energy efficiency, migration and computation costs. For example, if half of the switches are SDN and half of the nodes execute VNFs, the proposed algorithm can save up to 50\%, 40\% and 40\% power in small size, medium size and large size networks compared to non SDN/NFV networks, respectively, while the computation and migration costs are significantly reduced compared to the full SDN/NFV networks.
Due to the complexity of the energy efficiency problem in SDN and NFV networks, this paper does not consider traffic dynamics. Which are left for future work.

It should be noted that in this paper, only the energy consumption of SDN and NFV networks were investigated, and based on that, we concluded that partial SDN and hybrid NFV networks had very close energy efficiency performance to those achieved by full SDN and NFV networks in some cases. Because of its vast advantages, encompassing CAPEX and OPEX reduction, network flexibility,  QoS, easier network management and orchestration, full SDN, and full NFV are going to be the only choice in the future.

Jointly optimizing this parameter and the physical location of NFV, SDN, and traditional nodes to attain the best performance of network in terms of energy efficiency and related costs are considered as future works of this paper.

\bibliographystyle{ieeetr}

\end{document}